\documentclass[11pt]{myclass}
\newcommand{\bE}{{\mathbf E}}
\newcommand{\frmsd}{\textsc{frmsd}\xspace}
\newcommand{\rmsd}{\textsc{rmsd}\xspace}

\title {Outlier Robust ICP for Minimizing Fractional RMSD}
\author       {Jeff M. Phillips \;\;\; Ran Liu \;\;\; Carlo Tomasi
        \\
	Duke University, Durham, NC, USA 
	\\
	Duke University Technical Report:  CS-2006-05
       }

\begin{document}

\maketitle

\begin{abstract}
We describe a variation of the iterative closest point (ICP) algorithm for aligning two point sets under a set of transformations.  Our algorithm is superior to previous algorithms because (1) in determining the optimal alignment, it identifies and discards likely outliers in a statistically robust manner, and (2) it is guaranteed to converge to a locally optimal solution.  To this end, we formalize a new distance measure, fractional root mean squared distance (\frmsd), which incorporates the fraction of inliers into the distance function.  We lay out a specific implementation, but our framework can easily incorporate most techniques and heuristics from modern registration algorithms.  We experimentally validate our algorithm against previous techniques on 2 and 3 dimensional data exposed to a variety of outlier types.   
\end{abstract}

\section{Introduction}
Aligning an input data set to a model data set is fundamental to many important problems such as scanned model reconstruction~\cite{LPCRKPGADGSF00}, structural biochemistry~\cite{WSHB98}, and medical imaging~\cite{GKJB99}.  The input data and the model data are typically given as a set of points.  A point set may arise from laser scans of a 3D or 2D model, coordinates of atoms in a protein, positions of a lesions from a medical patient, or some other sparse representation of data.  However, the relative positions of these point sets is not known, making the task of registering them nontrivial.  

A popular approach to solving this problem is known as the iterative closest point (ICP) algorithm~\cite{BM92, CM92} which alternates between finding the optimal correspondence between points, and finding the optimal transformation of one point set onto the other.  As both steps reduce the distance between the point sets, this process converges, but only to a local minimum.  The effectiveness, simplicity, and generality of this algorithm has led to many variations~\cite{Zha94, RL01, PHYH04, CSK05, DF02, GIRL03, MGPG04,  WSHB98}.  For instance, the set of legal transformations can be just translations, all rigid motions, or all affine transformations.  Other versions replace the optimal correspondence between points by aligning each data point to the closest point on an implicit surface of the model data~\cite{CM92}.  Or the traditional squared distance can be replaced with a more efficient and stable approximation to the squared distance function~\cite{LPZ03}.  A now slightly outdated, but excellent survey \cite{RL01} evaluates many of these techniques.  

Yet, because ICP only converges to a local minimum, there has been considerable work on expanding and stabilizing the funnel of convergence\textemdash the set of initial positions for which ICP converges to the correct local minimum.  Others have attempted to solve the global registration problem~\cite{LG05,GMGP05}, where for any initial alignment they attempt to find the optimal alignment between two point sets.  This is often done in two steps.  First find a rough global alignment by corresponding certain distinguishable feature points.  Second refine the alignment with ICP.

However, all of these algorithms are vulnerable to point sets with outliers.  Outliers may result from:
\begin{itemize}
\item measurement error,
\item spurious data that was ignored or missed in the model,
\item partial matches because the point sets represent overlapping, but not identical pieces of the same object,
\item interesting changes in the underlying object between time steps or among comparable objects.
\end{itemize}
In short, outliers are unavoidable.  
Because ICP will find correspondences for all points, and then find the optimal transformation for the entire point set, the outliers will skew the alignment.  Many heuristics have been suggested~\cite{DF02, CSK05} including only aligning points within a set threshold~\cite{Zha94,TL94}, but most of these techniques are not guaranteed to converge, and thus can possibly go into an infinite loop, or require an expensive check to prevent this.  If the fraction $f$ of points which are outliers is known, then Trimmed ICP~\cite{CSK05} can be used to find the optimal alignment of the most relevant fraction $f$ of points.  This algorithm is explained in detail in Section \ref{ssec:TrICP}.  However, this fraction is rarely known a priori.  If an alignment is given then RANSAC type methods~\cite{FB81,CHC99} can be used to determine a good threshold for determining these outliers.  There are also many ad hoc solutions to this problem.  However, if the outliers are excluded from the data set in a particular alignment, then the alignment is no longer optimal, since those outliers which were removed influenced how the points were initially aligned.  

\subsection{Our Contributions}
Our solution to these problems is to incorporate the fraction of points which are outliers into the problem statement and into the function being optimized.  To this end, this paper makes the following contributions: 
\begin{itemize}
\item We formalize a new distance measure between point sets which accounts for outliers:  \frmsd.  This definition extends the standard \rmsd to account for outliers (Section \ref{sec:frmsd}).
\item We provide an algorithm, Fractional ICP, to optimize \frmsd (Section \ref{sec:ficp}) which we prove to converge to a local optimum in the correspondence, transformation, and fraction of outliers (Section \ref{sec:conv}).  
\item We give mathematical intuition for why \frmsd aligns data points which are more likely to be inliers than outliers (Section \ref{sec:just} and Section \ref{sec:lambda}).  
\item Finally, we empirically demonstrate that Fractional ICP identifies the correct alignment while simultaneously determining the outliers on several data sets (Section \ref{sec:exp}).
\end{itemize}

\section{Fractional RMS Distance}
\label{sec:frmsd}

Consider two point sets $D, M \in \mathbb{R}^d$.  The goal of this paper is to align an input data set $D$ to a model data set $M$ under some class of transformations, $\mathcal{T}$.  These may include rotations, translations, scaling, or all affine transforms.  We assume that these point sets are quite similar and there exists a strong correspondence between most points in the data.  There may, however, be outliers, points in either set which are not close to any point in the other set.  Our goal is to define and minimize over a set of transformations a relevant distance between these two point sets.  To aid in this, we define a matching function $\mu : D \to M$, which unless defined otherwise or given as a parameter, simply matches each point of $D$ to the closest point in $M$.  

\begin{defn} 
\textbf{[\rmsd] }
The root mean squared distance (or \rmsd) between two point sets $D,M \subset \mathbb{R}^d$, for a given matching $\mu: D \to M$ is the square root of the average squared distance between matched points:
$$\rmsd(D,M, \mu) = \sqrt{\frac{1}{|D|} \sum_{p \in D} ||p - \mu(p)||^2}$$
\label{def:rmsd}
\end{defn}

When convenient we sometimes write $\rmsd(D,M)$, letting $\mu$ match every point in $D$ to the closest point in $M$.  

\begin{problem}
\label{prb:rmsd}
\textbf{[minimizing \rmsd]}
Given a model point set $M$ and an input data point set $D$ where $D, M \subset \mathbb{R}^d$, compute the transformation $T \in \mathcal{T}$ to minimize $\rmsd(T(D),M)$:
$$\min_{\begin{array}{c} T \in \mathcal{T} \end{array}} \sqrt{\frac{1}{|D|}\sum_{p \in D} ||T(p)-\mu(p)||^2}.$$
\end{problem}

Problem \ref{prb:rmsd} is algorithmically difficult because as $T$ varies, so does the optimal matching $\mu$.  Also, \rmsd is quite susceptible to outliers because the squared distance gives a large weight to outliers.  To counteract this, a specific fraction $f \in [0,1]$ of points from $D$ can be used in the alignment and in the distance measure between the point sets.  These $f  |D|$ points can be chosen to solve Problem \ref{prb:rmsd} by selecting the points which have the smallest residual distance $r = ||p - \mu(p)||$.  Let $D_f = \{p \in D \mid |D_f| = \lfloor f |D|\rfloor \textrm{ and } \rmsd(D_f, M) \textrm{ is minimized} \}$.  But what fraction of points should be used?  We can always make $\rmsd(D_f,M) = 0$ by setting $f=1/|D|$ and aligning any single point exactly to another point.  So \rmsd by itself is no longer a viable measure.  For this reason, we propose a new distance measure.

\begin{defn} 
\textbf{[\frmsd] }
The fractional root mean squared distance (or \frmsd) is defined as follows:
$$\frmsd(D,M,f,\mu) = \frac{1}{f^\lambda} \sqrt{\frac{1}{|D_f|} \sum_{p \in D_f} ||p - \mu(p)||^2}$$
\label{def:frmsd}
\end{defn}

We will empirically and mathematically justify a value of $\lambda$ in Section \ref{sec:val-lambda} and Section \ref{sec:lambda}.  Again, it is sometimes convenient to let $\frmsd(D,M,f) = \frmsd(D,M,f,\mu)$ because $\mu$ can still be determined by $D$ and $M$.  This leads to a new, more relevant problem.  

\begin{problem}
\textbf{[minimize \frmsd]}
Given a model point set $M$ and an input data point set $D$ where $D, M \subset \mathbb{R}^d$, compute the transformation $T \in \mathcal{T}$ and fraction $f \in [0,1]$ to minimize $\frmsd(T(D),M,f)$:
$$\min_{\begin{array}{c} T \in \mathcal{T} \\ f \in [0,1] \end{array}} \frac{1}{f^\lambda} \sqrt{\frac{1}{|D_f|} \sum_{p \in D_f} ||T(p) - \mu(p)||^2}.$$
\label{prb:fRMSD}
\end{problem}

Intuitively, the $\frac{1}{f^\lambda}$ term serves to balance the \rmsd term.  $\frac{1}{f^\lambda}$ goes to $\infty$ as $f$ goes to $0$, while the \rmsd goes to $0$ as $f$ goes to $0$.  \frmsd, unlike \rmsd over any fraction of the data points, cannot equal $0$ unless some fraction of points align exactly.  Of course, one point can always align exactly to another point in the other subset, so in the implementation we restrict that $f > 1/|D|$, although this case is degenerate and almost never happens in practice.  Some arbitrary nonzero minimum value of $f$ can be set as desired.

\section{Algorithms}
In this section we describe algorithms to solve Problem \ref{prb:fRMSD}.  

\subsection{Trimmed ICP}
\label{ssec:TrICP}
The Trimmed ICP algorithm \ref{alg:TrICP} assumes $f$ to be given and computes a transformation $T \in \mathcal{T}$ of a point set $D$ to minimize \rmsd between $D_f$ and a model point set $M$.  When $f = 1$, this is the ICP algorithm~\cite{BM92}.  The algorithm iterates between computing the optimal matching $\mu$ and the optimal transform $T$ over the $f |D|$ closest points.  This algorithm has been shown~\cite{CSK05} to converge to a local minimum of $\rmsd(D_f, M)$ over all rotations, translations, and matchings.

\begin{algorithm}[h!!t]
\label{alg:TrICP}
\caption{TrICP$(D,M,f)$}
\begin{algorithmic}[1]
\STATE Compute $\mu_0 = \displaystyle{\arg\min_{\mu_0 : D \to M}} \rmsd(D,M, \mu_0)$.  
\STATE $i \leftarrow 0$.
\REPEAT  
\STATE $ i \leftarrow i+1$.
\STATE Compute $D_f $ minimizing $\rmsd(D_f, M)$ such that $D_f \subseteq D$ and $|D_f| = \lfloor f |D| \rfloor$.  
\STATE Compute $T \in \mathcal{T}$ minimizing $\rmsd(D_f,M)$.  \hspace{.3in} $D \leftarrow T(D)$.
\STATE Compute $\mu_i : D \to M$ minimizing $\rmsd(D,M)$.
\UNTIL {$(\mu_i = \mu_{i-1})$}
\end{algorithmic}
\end{algorithm}

In practice, the comparison on line 8 of Algorithm \ref{alg:TrICP}, $(\mu_i = \mu_{i-1}$), can be replaced by checking whether the $\rmsd(D,M)$ value decreases by less than some threshold at each step.  TrICP, however, does not completely solve Problem \ref{prb:fRMSD}; $\frmsd(D,M)$ is not minimized with respect to $f$.  It has been suggested~\cite{CSK05} to run TrICP for several values of $f$.  In fact, those same authors hypothesize that the $\frmsd(D,M,f)$ values returned from TrICP$(D,M,f)$ are convex in $f$, allowing them to perform a golden ratio search technique to avoid checking all values of $f$.  This hypothesis is easily shown to be false.  Also this technique fails to guarantee that the solution is a local minimum in the space of all transformations, matchings, and fractions.  The value attained by TrICP depends on the initial position of $D$ and $M$.  Thus, for the transformation $T$ calculated by TrICP, potentially another fraction $f$ can give a smaller value of $\rmsd(D_f, M)$ or of $\frmsd(D, M, f)$.  




\subsection{Fractional ICP}
\label{sec:ficp}
A simple modification of TrICP, shown in Algorithm \ref{alg:FICP}, will actually provide the desired local minimum.  We refer to this algorithm as Fractional ICP or FICP.

\begin{algorithm}[h!!t]
\label{alg:FICP}
\caption{FICP$(D,M)$}
\begin{algorithmic}[1]
\STATE Compute $\mu_0 = \displaystyle{\arg\min_{\mu_0 : D \to M}} \rmsd(D,M, \mu_0)$.  
\STATE Compute $f_0 \in [0,1]$ minimizing $\frmsd(D,M,f_0,\mu_0)$.  
\STATE $i \leftarrow 0$.
\REPEAT
\STATE $ i \leftarrow i+1$.
\STATE Compute $D_f $ minimizing $\rmsd(D_f, M)$ such that $D_f \subseteq D$ and $|D_f| = \lfloor f |D| \rfloor$.  
\STATE Compute $T \in \mathcal{T}$ minimizing $\rmsd(D_f,M)$.  \hspace{.3in} $D \leftarrow T(D)$.
\STATE Compute $\mu_i : D \to M$ minimizing $\rmsd(D,M, \mu_i)$.
\STATE Compute $f_i \in [0,1]$ minimizing $\frmsd(D,M,f_i,\mu_i)$.
\UNTIL {$(u_i = u_{i-1}$ and $f_i = f_{i-1})$}
\end{algorithmic}
\end{algorithm}

Again, in practice, the comparison on line 10 of Algorithm \ref{alg:FICP} can be replaced be checking whether the $\frmsd(D,M,f)$ value decreases by less than some threshold at each step.

\subsection{Implementation}

To implement TrICP we need 3 operations: computing the matching, computing the subset $D_f$, and computing the transformation.  To implement FICP we need the additional step of computing the fraction.  

\subsubsection{Computing the Matching}
For each point $p \in D$ we need to find its closest point $m \in M$.  Since $M$ is fixed through the algorithm, we can precompute a hierarchical data structure which can quickly return the nearest neighbor.  We implemented a $kd$-tree, at a one-time, initial cost of $O(|M| \log |M|)$.  The nearest neighbor can be returned in $O(\log |M|)$ time.  This operation is required for each point in $|D|$.  So the matching can be computed in $O(|D| \log |M|)$.  This is in general the most time consuming step of the algorithm.

We could replace the $kd$-tree with a $d^2$-tree~\cite{LPZ03}, or when appropriate use point to surface matching as in \cite{CM92} or \cite{RL01}, but we would loose our guarantee of convergence.

\subsubsection{Computing the Subset $D_f$}
The set $D_f = \{p \in D \mid |D_f| = \lfloor f |D| \rfloor, \rmsd(D_f, M) \textrm{ is minimized} \}$ is defined by the $\lfloor f |D| \rfloor$ points with the smallest residual distances $r = ||p - \mu(p)||$.  This observation implies the following algorithm.  Compute and sort all residual distances and let $D_f$ be the $f |D|$ points with the smallest residual distances.  The runtime is bounded by the sorting which takes $O(|D| \log |D|)$ time.  

\subsubsection{Computing the Transformation}
\label{sec:compT}
The set of allowable transformations, $\mathcal{T}$, may include rotations, translations, and scalings.  Or it may be as general as all affine transformations.  When we consider rotations, translations, and scalings, Problem \ref{prb:rmsd} is written:
$$\min_{\begin{array}{c} R \in SO(d) \\ t \in \mathbb{R}^d \\ s \in \mathbb{R} \end{array}} \sqrt{\frac{1}{|D|} \sum_{p \in D} ||s R(p) + t - \mu(p)||^2}.$$
For a fixed matching, $\mu$, this is known as the absolute orientation problem, and can be solved exactly~\cite{Hor87} in $O(n^2)$ time.  When $d \leq 3$, this can be solved in $O(n)$ time~\cite{WSV91}.   There are actually 4 distinct algorithms\textemdash one using rotation matrices and the SVD~\cite{HN81}, one using rotation matrices and the eigenvalue decomposition~\cite{SS87}, one using unit quaternions~\cite{FH83}, and one using dual number quaternions~\cite{WSV91}\textemdash but all are in practice approximately equivalent in run time~\cite{ELF97}.  We use the simplest technique~\cite{HN81} which reduces the solution to computing an SVD.  

When $\mathcal{T}$ is the set of all affine transformations, Problem \ref{prb:rmsd} is written:
$$\min_{\begin{array}{c} A\in \mathcal{T} \end{array}} \sqrt{\frac{1}{|D|} \sum_{p \in D} ||A(p) - \mu(p)||^2},$$
where $A$ is an affine transformation.
This reduces to a generic least squares problem that can be solved with a matrix inversion.  

\subsubsection{Computing the Fraction}
\label{sec:fraction}
There are only $|D|$ fractions which we need to consider.  Consider the sorted order of the point set $D$ by each point's residual distance $r = ||p - \mu(p)||$.  Each prefix of this ordering represents a distinct fraction.  If we maintain the value $\sum_{p \in D_f} ||p - \mu(p)||^2$ for each $D_f$ we can compute $\frmsd(D,M,f)$ in constant time for a given fraction $f$.  We can also update $D_f$ to a point set of size $|D_f| + 1$ in constant time by adding the next point in the sorted order to $D_f$.  If the $i$th prefix yields the smallest value of \frmsd, then $f$ is set to $i / |D|$.  So this computation takes $O(|D|)$ time.

\section{Convergence of Algorithm}
\label{sec:conv}

In this section we show that Algorithm \ref{alg:FICP} converges to a local minimum of $\frmsd(D,M,f)$ over the space of all transformations, matchings, and fractions of points used in the matching.  This is a local minimum in a sense that if all but one of transformations, matchings, or fractions is fixed, then the value of the remaining variable cannot be changed to decrease the value of $\frmsd(D,M,f)$.  

\begin{theorem}
For any two points sets $D,M \in \mathbb{R}^d$, Algorithm \ref{alg:FICP} converges to a local minimum of $\frmsd(T(D),M,f,\mu)$ over $(f, T, \mu) \in  [0,1] \times \mathcal{T} \times \{D \to M\}$.
\end{theorem}
\begin{proof}
Algorithm \ref{alg:FICP} only changes the value of $(f,T,\mu)$ when computing the optimal transformation $T$ (line 7), computing the optimal matching $\mu$ (line 8), or computing the optimal fraction $f$ (line 9).  None of these steps can increase the value of $\frmsd(D,M,f_i, \mu_i)$, because staying at the current value would retain the value of $\frmsd(D,M,f,\mu)$, but each can potentially decrease it.  (When two possible values of $(f,\mu,T)$ degenerately produce the same value of $\frmsd(D,M,f,\mu)$, we consistently choose the smaller one according to some consistent, but arbitrary ordering.)

FICP terminates only when $\mu_i = \mu_{i-1}$ and $f_i = f_{i-1}$.  The optimal transformation computed at iteration $i$ (line 7) is a function of the matching of the points $\mu_{i-1}$ and which points are included, which is determined by $f_{i-1}$.  Thus, the transformation will only change in iteration $i+1$ if $\mu_i$ or $f_i$ change from $\mu_{i-1}$ or $f_{i-1}$, respectively.  If $\mu_i = \mu_{i-1}$ and $f_i = f_{i-1}$ then FICP will terminate, and $(f_i,T,\mu_i)$ will be a local minimum.  If it were not, then either $f$ or $\mu$ would have changed in the last iteration, and $\frmsd(D,M,\mu,f)$ would have decreased or stayed the same in the $i$th iteration.

Furthermore, FICP terminates in a finite number of iterations, because there are $|D|$ possible values of $f$ and $|M|^{|D|}$ possible values of $\mu$, and the algorithm can never be at any of these locations twice.  
\end{proof}

In practice the convergence is much faster than the upper bound of $|D| \cdot |M|^{|D|}$ steps.  ICP has recently~\cite{ESE05} been shown to require $\Omega(|D \cup M| \log |D \cup M|)$ iterations for certain adversarial inputs; however, these rarely occur in practice.  Furthermore, Pottmann \emph{et. al.}~\cite{PHYH04}, have shown that ICP has linear convergence when it is close to the optimal solution and a point-to-point matching is used.  However, ICP has quadratic convergence when using a point-to-surface or other similar matching criterion as described in~\cite{PHYH04} or~\cite{MGPG04}.  The lower bounds clearly hold for TrICP.  The upper bounds, in terms of convergence rates, intuitively hold, but the reduction seems a little more complicated.  Such a proof is outside the scope of this paper.  

\section{Data Generation Model}
\label{sec:just}
In order to formalize the expected mathematical properties of the \frmsd measure and the FICP algorithm, we now state some fairly general assumptions about the input data.  All data on which FICP is used need not these exact properties, but we hope that these properties are general enough that whatever differences exist in the alternative data will not significantly affect the following analysis and the resulting conclusions.

Since data come from a measurement process that might generate spurious
measurements as well as miss valid ones, we do not require every data point to
have a corresponding model point, or viceversa. Specifically, we assume that
data points are generated from model points by the following abstract
procedure:
\begin{enumerate}
\item Generate a set $M_{I}$ of model points that will have
corresponding data points (the subscript $I$ stands for
``inlier'').
\item For every model point $m\in M_{I}$, let
\[
p = T^{-1}(m + n)
\]
be the corresponding data point, where $T$ is a transformation in
the set ${\mathcal T}$ and $n$ is isotropic Gaussian noise with
standard deviation $\sigma$. The set of data points $p$
corresponding to $M_{I}$ is denoted as
$D_{I}$.\label{step:inliers}
\item Generate a random set $D_{O}$ of data outliers.
\item Generate a random set $M_{O}$ of model outliers out of a spatial Poisson
process.
\end{enumerate}
We let $D = D_{I} \cup D_{O}$ and $M = M_{I} \cup M_{O}$. Let
$p_{I}$ be the fraction of data inliers relative to all data
points. The detailed spatial statistics of data outliers are
irrelevant to our analysis. The Poisson process for model outliers
is a minimally informative prior. We let the density of this
process be $\omega$ points per unit volume.

The probability density of the squared magnitude $z = \|n\|^{2}$
of the correspondence noise is a chi square density in $d$
dimensions:
\[
g_{\chi^{2}(d)}(z) =
\frac{z^{d/2-1}}{2^{d/2}\sigma^{d}\Gamma(d/2)}\,e^{-\frac{z}{2\sigma^{2}}}
\]
where
\[
\Gamma(x) = \int_{0}^{\infty}\, t^{x-1}\, e^{-t}\, dt
\]
is the gamma function. 
In particular,
\begin{eqnarray*}
\Gamma(0) &=& 0\\
\Gamma(1) &=& 1\\
\Gamma(n) &=& (n - 1)! \hspace*{0.5cm} \mbox{ for integer $n > 1$}\\
\Gamma(1/2) &=& \sqrt{\pi} \;\approx\; 1.77245\\
\Gamma(n + 1/2) &=& \sqrt{\pi} \frac{1\cdot 3 \cdot 5 \cdot \ldots
\cdot (2n - 1)}{2^{n}} \hspace*{0.5cm} \mbox{ for integer $n >
0$.}
\end{eqnarray*}

The expected number of model outliers in a region of space with volume $V$ is
equal to $\omega V$.

Suppose now that the correct geometric transformation $T\in {\mathcal T} $ is
applied to data point $p$ to obtain the transformed data point
\[
q = T(p) = m + n
\]
(see step \ref{step:inliers} in the data generation model above).

If $q$ and $m$ correspond, their distance statistics are chi square. If $q$ and
$m$ do not correspond, the situation is more complex: Either point (or both)
could be an outlier, or they could be non-corresponding inliers. We do not know
the distance statistics for model inliers. In the remainder of this section, we
assume that the probability that 
a data inlier is nearest to a non-corresponding model inlier is
negligible. Under this assumption, the probability density of the distance $r$
from $q$ to the nearest outlier, given that model outliers are from a spatial
Poisson process with density $\omega$ points per unit volume, can be shown to
be
\[
w(r) = \omega\,S(d)\,r^{d-1}\,e^{-\omega\,S(d)\,r^{d} / d}
\hspace*{0.5cm}\mbox{ for $r \geq 0$}
\]
where
\[
S(d) = \frac{2\pi^{d/2}}{\Gamma(d/2)}
\]
is the surface of the unit sphere in $d$ dimensions and $\Gamma(\cdot)$ is the gamma function. The function $w(r)$ is
known as the Weibull density with shape parameter $d$ (equal to
the dimension of space) and scale parameter
\[
s(d,\, \omega) = \frac{1}{\sqrt{d}}\;
\sqrt[d]{\frac{d\,\Gamma(d/2)}{2\omega}}\;.
\]

So far we have not specified the units of measure. Since $\sigma$ is a distance
and $\omega$ is a distance raised to power $-d$ (density per unit volume), the
parameter $\sigma\omega^{1/d}$ is dimensionless. As long as $\sigma$ and
$\omega$ are properly scaled to each other, the analysis that follows is
independent of $\sigma$.

\section{The Value of $\lambda$}
\label{sec:lambda}

In this Section we justify a particular choice for the value of $\lambda$ used
in the definition of the fractional root mean squared distance (\frmsd).

As shown in Section \ref{sec:fraction}, the FICP
algorithm selects a fraction $f$ of data-model matches in increasing order of
their residual distances $r = \|p - \mu(p)\|$ between data points $p$ and their
nearest model points $\mu(p)$. Because of this, choosing a fraction $f$ is
equivalent to choosing a maximum allowed value $r^{*}$ for the residual
distance $r$. Since we would like the FICP algorithm to favor inliers over
outliers, it makes sense to require $r^*$ to be defined in such a way that data
points that are $r^*$ away from a model point are equally likely to be inliers
as they are to be outliers. Let us call such a value of $r^*$ the {\em critical
distance}. We then ask the following question: {\em Is there a value of
$\lambda$ in the definition of the \frmsd for which the value of $f$ chosen by
the FICP algorithm corresponds to the critical distance?}

To answer this question, we first express $r^{*}$ as a function of the model
parameters (Section \ref{subsec:critical}), and determine the function that
relates an arbitrary distance $r$ to the corresponding fraction $f$ (Section
\ref{subsec:f}). We then write an estimate of the \frmsd under an ergodicity
assumption (Section \ref{subsec:FRMSDest}). This estimate is itself a function
of $f$, and therefore of $r$. The FICP algorithm maximizes the \frmsd with
respect to $f$, that is, finds a zero for the derivative of the \frmsd with
respect to $f$. Setting the value of $f$ where this zero is achieved to
$f(r^{*})$ yields an equation for $\lambda$, whose solution set justifies our
choice for this parameter (Section \ref{subsec:lambda}).

Our analysis holds for outlier densities $\omega$ that are below a
certain value $\omega_{\max}$, which is inversely proportional to
the standard deviation $\sigma$ of the noise that affects the data
points. If outliers exceed this density, then matching data and
model points based on minimum distance is too unreliable to yield
good results.

\subsection{The Critical Distance}\label{subsec:critical}

Define $r^{*}$ so that a data and a model point at distance $r^{*}$ from each
other are equally likely to correspond to each other as they are not to. This
section derives an expression for $r^{*}$ as a function of the standard deviation $\sigma$ of the correspondence noise, the density $\omega$ of the spatial Poisson process that generates unmatched points, and the dimension $d$ of space.

The volume of a sphere of radius $r$ in $d$ dimensions is 
$$V_s(r) = \frac{S(d)}{d} r^d$$
where $S(d)$ was defined in Section \ref{sec:just}.  
The volume of
the shell between radii $r$ and $r+\delta r$ is
\[
\delta V_{s} =  V_s(r+\delta r) - V_s(r) = \frac{S(d)}{d} \,\left[ (r +
\delta r)^{d} - r ^{d}\right] \approx S(d) \, r^{d-1}\, \delta r \;.
\]
This approximation is asymptotically exact as $\delta r \rightarrow
0$.

The probability mass in the same shell for
an isotropic Gaussian distribution with zero mean and standard
deviation $\sigma$ is
\[
\delta G_{s} = 2r\, g_{\chi^{2}(d)}(r^{2})\, \delta r =
\frac{S(d)}{(2\pi)^{d/2}\,\sigma}\;
\left(\frac{r}{\sigma}\right)^{d-1}\;
e^{-\frac{1}{2}\,\left(\frac{r}{\sigma}\right)^{2}}\; \delta r
\]
as $\delta r\rightarrow 0$ 
(the term $2r$ derives from the Jacobian of the transformation $z = r^{2}$, since the $\chi^{2}$ density is defined for the square of a distance)
.

Assume that the center of the shell above is at the transformed data point $q$
defined in Section \ref{sec:just}. As explained in Section \ref{sec:just}, if
$q$ and $m$ correspond, their distance statistics are chi squared, and the
likelihood of a particular radius $r$ is $\delta G_{s}/ \delta r$. Otherwise, the distance
statistics are approximately described by a spatial Poisson process with
density $\omega$. Then, the critical distance is determined by the equation
\[
\omega \, \delta V_{s} = \delta G_{s}
\]
that is,
$$\omega S(d) r^{d-1} \delta r = \frac{S(d)}{(2\pi)^{d/2} \sigma} \left(\frac{r}{\sigma}\right)^{d-1} e^{-\frac{1}{2} \left(\frac{r}{\sigma}\right)^2} \delta r$$
which can be simplified to the following:
\begin{equation}
e^{-\frac{1}{2}\,\left(\frac{r}{\sigma}\right)^{2}} \;=\; \omega\,
\sigma^{d}\, (2\pi)^{d/2}\;. \label{eq-crossing}
\end{equation}

The left-hand side of equation (\ref{eq-crossing}) is strictly positive and
monotonically decreasing in $r$ and the right-hand side is constant, so the
equation admits a solution if and only if
\[
0 < \omega \leq \omega_{\max} =
\frac{1}{(\sqrt{2\pi}\,\sigma)^{d}}\;.
\]
If the outliers exceed this maximum density $\omega_{\max}$, the critical
distance shrinks to zero: any model point around any given data point $q$ is
more likely to be an outlier than it is to be the model point corresponding to
$q$. Of course, when there are no model outliers ($\omega = 0$) the concept of
critical distance loses its significance.

\begin{figure}[ht]
\centerline{ \includegraphics[width=8cm]{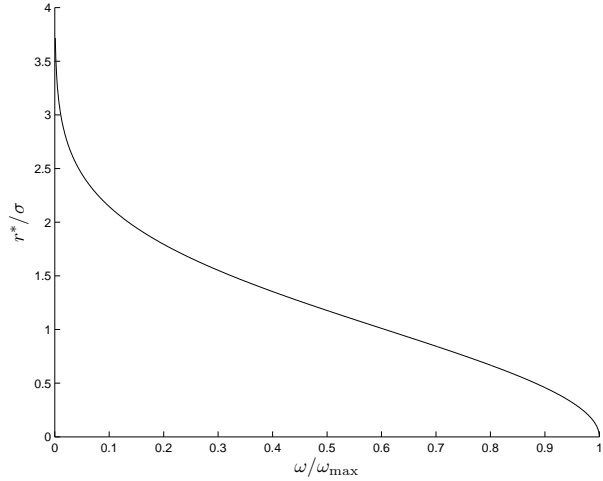} } \caption{Critical distance
normalized by noise standard deviation, plotted versus model outlier density
normalized by maximum density. } \label{fig:rho}
\end{figure}

Equation (\ref{eq-crossing}) can be solved for $r$ to yield the
desired value of $r^{*}$ as a function of the model parameters:
\[
\frac{r^{*}}{\sigma} = \sqrt{-2\,\log_{e}((\sqrt{2\pi}\,
\sigma)^{d}\omega)} = \sqrt{2\,\log_{e}\,
\frac{\omega_{\max}}{\omega}}\;.
\]
The critical distance normalized by $\sigma$ and expressed as a
function of $\alpha = \omega / \omega_{\max}$ is
\[
\rho(\alpha) = \frac{r^{*}(\alpha)}{\sigma} = \sqrt{-2\,
\log_{e}\alpha}\;.
\]
This function is independent of all model parameters and is
plotted in Figure \ref{fig:rho}.

\subsection{Relationship between $f$ and $r$}\label{subsec:f}

As explained earlier, to every fraction $f$ of data points
considered by the FICP algorithm there corresponds a maximum
distance $r$, in the sense that $f|D|$ data-model point pairs have
distance at most $r$. Consider a particular data point $p$ and its
transformed version $q = T(p)$. If the data generation process is
ergodic, the fraction $f$ equals the probability that the nearest
model point $m$ to a point $q$ selected at random from the
transformed data set $T(D)$ is at most $r$ away.

With probability $p_{I}$, the data point $q$ has a corresponding
model point (inlier). In this event, if $r_{I}$ is the distance
from this model point and $r_{O}$ is the distance from the nearest
model outlier point, the complement of the cumulative probability
function of the distance $r$ to the nearest model point (either
inlier or outlier) is
\begin{eqnarray*}
\lefteqn{1 - F(r) = 1 - {\mathcal P}[\min(r_{I}, r_{O}) < r] = {\mathcal
P}[\min(r_{I}, r_{O}) \geq r]}\\ &=& {\mathcal P}[r_{I} \geq r
\;\cap\; r_{O} \geq r] = {\mathcal P}[r_{I} \geq r] \; {\mathcal P}[r_{O}
\geq r]\\ &=& (1 - {\mathcal P}[r_{I} \leq r]) \; (1 - {\mathcal P}[r_{O}
\leq r]) = (1 - F_{I}(r)) \; (1 - F_{O}(r))
\end{eqnarray*}
where $F_{I}(r)$ and $F_{O}(r)$ are respectively the probability
that the matching model point and the nearest model outlier are at
most $r$ units away from $q$. From Section \ref{sec:just}, these
probabilities are as follows:
\[
F_{I}(r) = \int_{0}^{r^{2}}\; g_{\chi^{2}(d)}(\zeta)\, d\zeta
\]
and
\[
F_{O}(r) = \int_{0}^{r}\; w(\rho)\, d\rho\;.
\]
Then, if $q$ has a corresponding model point, the density of its distance from
the nearest model point is
\begin{eqnarray*}
\phi_{c}(r) &=& \frac{dF(r)}{dr} = - \frac{d}{dr}(1 - F(r)) \\
&=& 2r\,g_{\chi^{2}(d)}(r^{2})\, (1 - F_{O}(r)) \;+\; (1 - F_{I}(r))\, w(r)\;.
\end{eqnarray*}

With probability $p_{O} = 1 - p_{I}$, the data point $q$ is
instead an outlier. Then, it has no corresponding model point, so
the probability that the nearest model point is at most $r$ units
away is simply $F_{O}(r)$. In summary, the probability density of
the distance between a data point $q$ and its nearest model point
$\mu(q)$ is
\[
\phi(r) = p_{I}\, \phi_{c}(r) + p_{O}\, w(r)
\]
and the average fraction of model points within $r$ units from a
data point is
\[
f(r) = \int_{0}^{r}\, \phi(\rho)\, d\rho = p_{I}\, \int_{0}^{r}\,
\phi_{c}(\rho)\, d\rho + p_{O}\, F_{O}(r)\;.
\]

The derivative of $f$ with respect to $r$ is $\phi(r)$.

\subsection{Ergodic Estimate of the \frmsd}\label{subsec:FRMSDest}

An estimate of the fractional root mean squared distance (\frmsd) can be
obtained by assuming ergodically that the sample moment included in the definition of \frmsd
is close to the corresponding statistical moment:
\[
\frac{1}{f|D|}\sum_{p\in D_{f}}\, \|p - \mu(p)\|^{2} \approx \bE_{p\in
D_{f}}[\|p - \mu(p)\|^{2}]\;.
\]
This assumption requires both ergodicity and a sufficient number $f |D|$ of data points that are close enough to the model points.  We can then write
\begin{eqnarray*}
\lefteqn{\frmsd^{2}(D, M, f) = \frac{1}{f^{2\lambda}}\;
\frac{1}{f|D|}\sum_{p\in D_{f}}\, \|p - \mu(p)\|^{2}}\\ &\approx&
\frac{1}{f^{2\lambda}}\; \bE_{p\in D_{f}}[\|p - \mu(p)\|^{2}] =
\frac{1}{f^{2\lambda}}\; \int_{0}^{r}\, \rho^{2}\, \phi(\rho)\,
d\rho \;.
\end{eqnarray*}

\subsection{Stationary Point of the \frmsd Estimate}\label{subsec:lambda}
\label{sec:stat-pnt}

At the minimum of $\frmsd(D, M, f)$, the derivative of $\frmsd^{2}(D, M, f)$ with
respect to $f$ is zero. Differentiation of the expression at the end of Section
\ref{subsec:FRMSDest} yields
\[
\frac{d}{df}\frmsd^{2}(D, M, f) \;=\; \frac{-2\lambda}{f^{2\lambda + 1}}\;
\int_{0}^{r}\, \rho^{2}\, \phi(\rho)\, d\rho \;+\; \frac{r^{2}}{f^{2\lambda}}\;
\phi(r)\, \frac{dr}{df}\;.
\]
Since
\[
\left(\frac{dr}{df}\right)^{-1} = \frac{df}{dr} = \phi(r)\;,
\]
the last addend simplifies to $r^{2} / f^{2\lambda}$, and
\[
f^{2\lambda}\, \frac{d}{df}\frmsd^{2}(D, M, f) = - \frac{2\lambda}{f}\;
\int_{0}^{r}\, \rho^{2}\, \phi(\rho)\, d\rho \;+\; r^{2}\;.
\]
Zeroing this derivative and setting $r = r^{*}$ and $f = f(r^{*})$ yields the
following equation for $\lambda$:
\[
\lambda = \frac{1}{2}\; \frac{(r^{*})^{2}\, \int_{0}^{r^{*}}\,
\phi(\rho)\, d\rho}{\int_{0}^{r^{*}}\, \rho^{2}\, \phi(\rho)\,
d\rho}\;.
\]

\begin{figure}[ht]
\centerline{
\includegraphics[width=8cm]{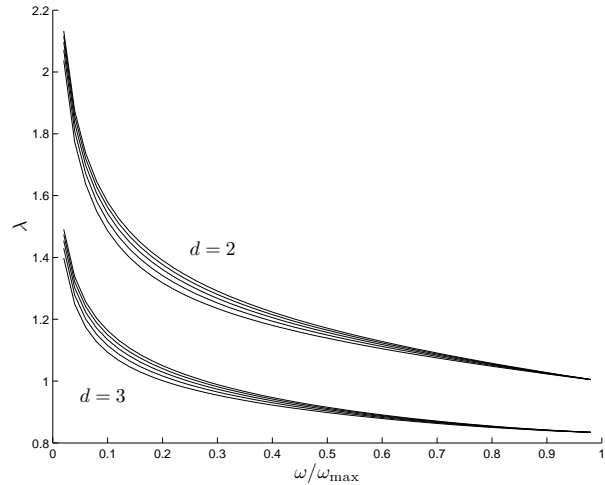}}
\caption{Theoretical value of $\lambda$ in the definition of the \frmsd in two
(upper bundle) and three (lower bundle) dimensions as a function of the
relative model outlier density $\omega/\omega_{\max}$ and for various values of
the data inlier fraction $p_{I}$. Curves in each bundle correspond to $p_{I} =
\{0.5, 0.6, 0.7, 0.8, 0.9\}$ from the bottom up. Dependency on $p_{I}$ is weak.}
\label{fig:lambda}
\end{figure}

Figure \ref{fig:lambda} plots the values of $\lambda$ in two and
three dimensions as a function of the relative model outlier
density $\omega/\omega_{\max}$ and for various values of the data
inlier fraction $p_{I}$.

Since the noise standard deviation $\sigma$ acts merely as an overall scale
factor, these plots do not depend on $\sigma$. It is apparent from the figure
that $\lambda$ depends weakly on the fraction $p_{I}$ of data inliers. The
knees of the plots are at about $\lambda = 1.3$ and $\lambda = 0.95$ for $d =
2$ and $d = 3$ dimensions, respectively, corresponding to $\omega /
\omega_{\max} = 0.2$. These knee values are selected as general-purpose values
for the definition of \frmsd in two and three dimensions.

\section{Experiments}
\label{sec:exp}
The main advantage of FICP over other variants of ICP is that it automatically determines the outlier set via a fraction $f$ and reaches a optimum in terms of the correspondence, the transformation, and the fraction of outliers.  In doing so, it takes less time than algorithms which have no guarantees, despite searching a larger parameter space.  We also demonstrate that the radius of convergence for FICP is expanded as compared to TrICP.  

Finally, we deal empirically with the issue of the parameter $\lambda$ used in the definition of \frmsd.  We observe that \frmsd is robust to the choice of $\lambda$ within a broad range.  However the radius of convergence and efficiency of FICP is improved when $\lambda$ is set to a slightly higher values than those determined optimal for identifying outliers in Section \ref{sec:lambda}.  Intuitively, a smaller value of $\lambda$ is more likely to classify correct correspondences as outliers when the alignment is not close, and thus get stuck in local minimum.  For higher values of $\lambda$ these types of local minimum seem less prevalent.
So for all performance studies we set $\lambda = 3$, unless otherwise specified.  For this value FICP has an expanded radius of convergence and tends to find very similar alignments as when $\lambda$ is set according to the analysis in Section \ref{sec:lambda}.   
After converging, we recommend setting $\lambda = 1.3$ for $d=2$ or $\lambda = .95$ for $d=3$ to identify outliers more agressively.  This final phase should take very few additional iterations of the algorithm, since, as we demonstrate, moderately modifying the value of $\lambda$ has small effects on the \frmsd and $f$ values returned.

\subsection{Data Sets}
We perform many tests on the SQUID fish contour database~\cite{SQUID} from the University of Surrey, UK.  This database has 1100 2D contours of fish and each contour has 500 to 3000 points.  The size of this data set allows us to average results over a very large set of experiments.  We do not know of any 3D database even close to this size, and it has been previously used to evaluate TrICP~\cite{CSK05}.  

We also perform some experiments on a limited number of 3D models.  In particular we use the \emph{bunny} and the \emph{happy Buddha} data set from the Stanford 3D Scanning Repository.  

\begin{figure}[htb]
  \centering
  \includegraphics[width=\linewidth]{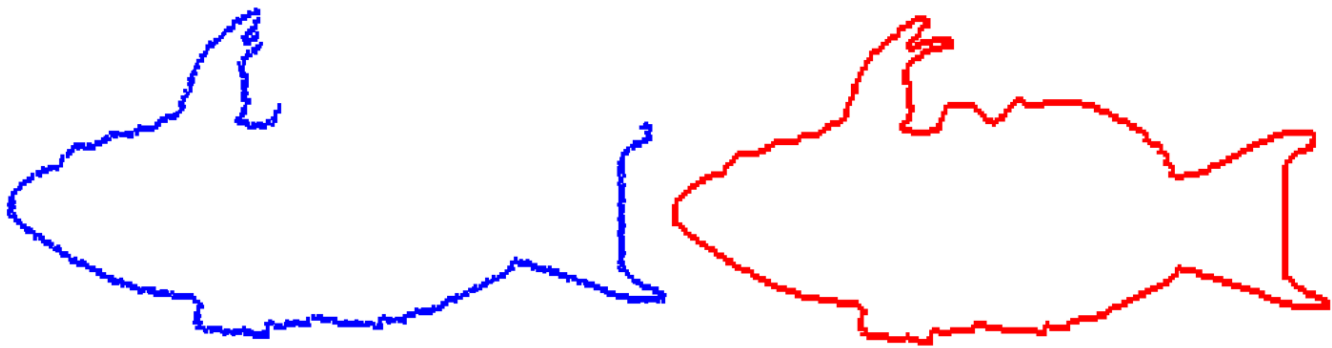}
  \caption{\label{fig:SQUID+O}
  	SQUID example with $M$ in blue suffering from \emph{Occlusion} noise (left), and $D$ in red (right).  $p_I = .75$}
\end{figure}

\begin{figure}[htb]
  \centering
  \includegraphics[width=\linewidth]{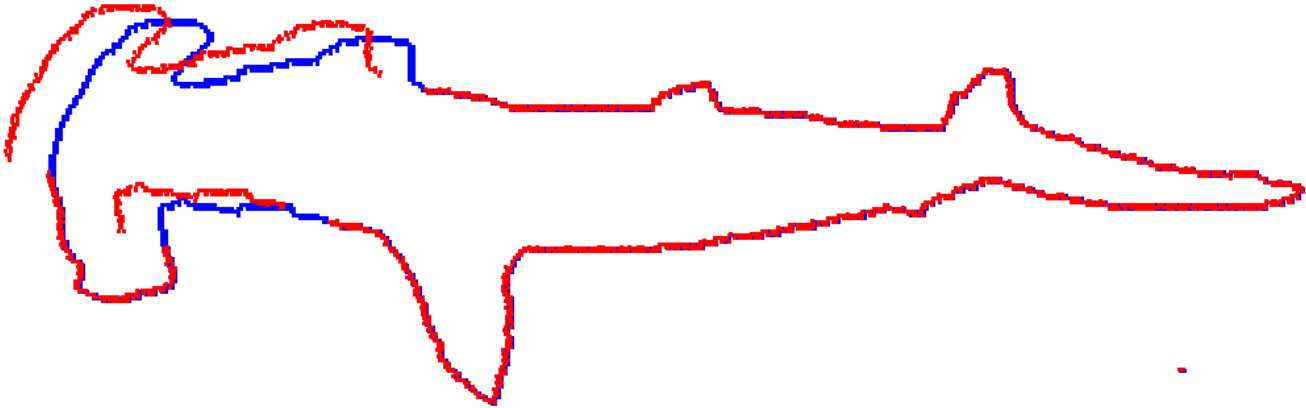}
  \caption{\label{fig:SQUID+D}
  	SQUID example with $M$ in blue superimposed on $D$ in red with \emph{Deformation} noise added.  $p_I = .75$}
\end{figure}

\begin{figure}[htb]
  \centering
  \includegraphics[width=.8\linewidth]{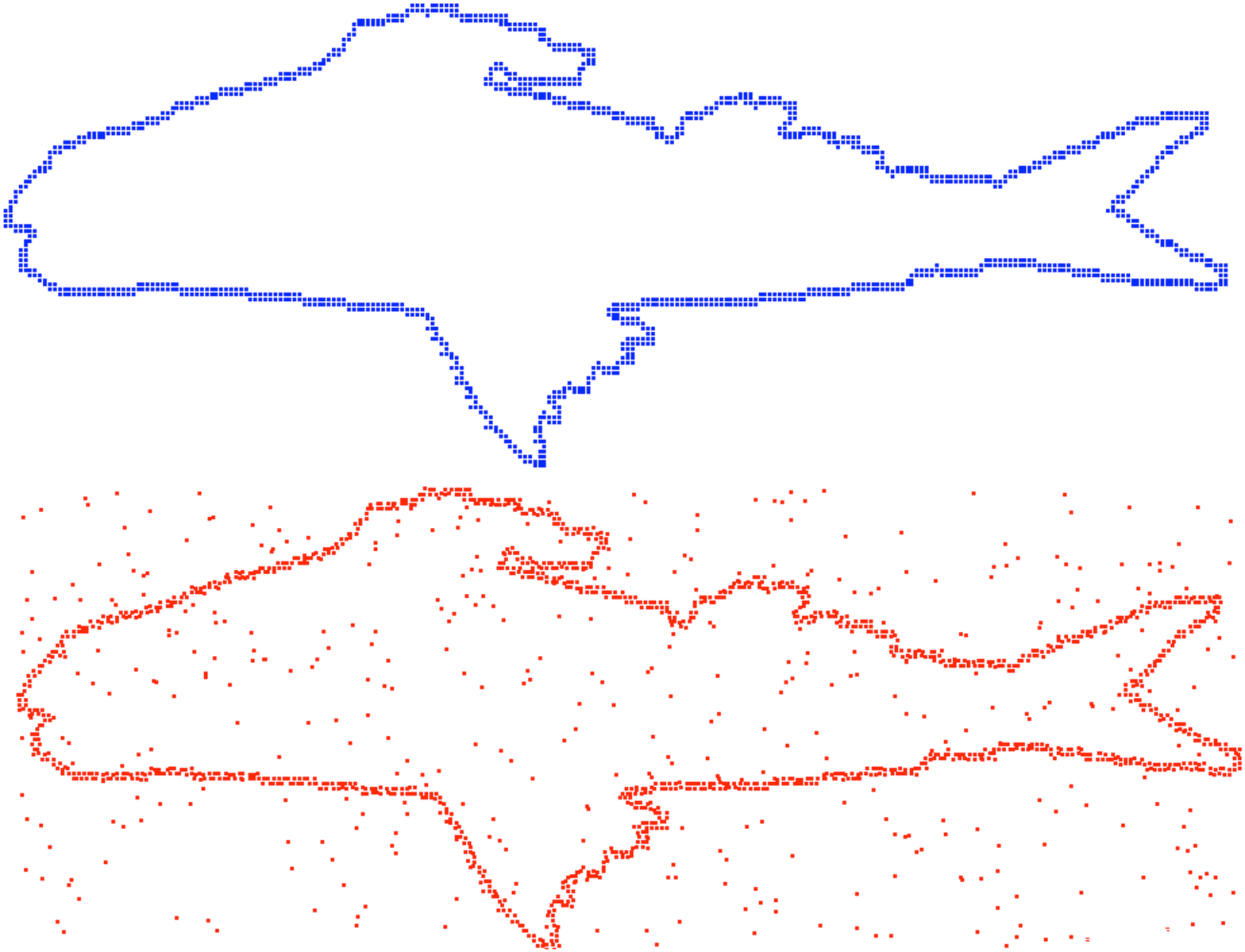}
  \caption{\label{fig:SQUID+N}
  	SQUID example with $M$ in blue (top), and $D$ with \emph{New Data} noise added in red (bottom). $p_I = .75$}
\end{figure}

We synthetically introduce outliers into the data sets in 3 ways.  We always begin by creating two copies $M$ and $D$, to represent the model and the input data, of the particular data set.  A parameter $p_I$ fraction of the final set $D$ are left undisturbed as data inliers.
\begin{itemize}
\item  \textbf{Occlusion:}  We randomly choose a ball $B$ and remove all of the points from $M$ within $B$.  This test represents cases where the model set is only partially observed because of occlusions, where there are two overlapping views of the same object that do not exactly align, or where the input data $D$ has grown since the model was formed.  An example is shown in Figure \ref{fig:SQUID+O}.
\item \textbf{Deformation:}  We randomly choose a ball $B$ and shift randomly the points $D \cap B$.  This represents the case where $D$ is deformed slightly between time steps.  See Figure \ref{fig:SQUID+D}.
\item  \textbf{New data:}  We add a set of points to $D$.  These points are placed uniformly at random within a bounding box of $D$.  This represents outliers caused by some sort of data retrieval noise or from spurious or new data.  See Figure \ref{fig:SQUID+N}.
\end{itemize}
Finally, we always introduce some further noise in the models.  For each point $p \in D$, we create a random vector $n$ distributed according to a Gaussian distribution with standard deviation $\sigma$, and we add $n$ to $p$.  

We perform many tests on synthetic data because we then know that a good match exists and it is thus easy to quantify the performance on our algorithm.

Additionally, we perform tests on real scanned data.  We align pairs of scanned images of the \emph{dragon} model from the Stanford 3D Scanning Repository from views $24^\circ$ or $48^\circ$ apart.  Because the different views observe different portions of the model, there are many points which have no good alignment in both the model and data set. These are outliers.

\subsection{Performance}
For each synthetic data set and type of outliers described above, we perform the following set of tests.  We first rotate $D$ by $\theta$ degrees where $\theta$ is from the set $\{5^\circ, 10^\circ, 25^\circ, 50^\circ\}$.  The axis of rotation is chosen randomly for the 3D data.  We then run ICP, TrICP searching for $f$ with the golden rectangle search, and FICP, minimizing over all rigid motions.  We report the total number of iterations of each, the run time, and the final values of \rmsd, \frmsd, and $f$.  We vary the input so that $p_I$ is either $\{.75, .88, .95\}$.  We expect that optimally $f$ should be near $p_I$ since in our data $\sigma$ is small compared to $\omega / \omega_{\max}$.  All experiments were performed on a 3 GHz Pentium IV processor with 1 Gb SD-RAM.

We show in Table \ref{tbl:SQUID-O5} the average performance of all algorithms on the entire SQUID data set where points are removed from $M$, giving $D$ occlusion outliers with $p_I = \{.75, .88, .95\}$.  Table \ref{tbl:SQUID-D5} shows the same where $D$ is given deformation outliers with $p_I = \{.75, .88, .95\}$.  Table \ref{tbl:SQUID-N5} shows where $D$ is given new data outliers with $p_I = \{.75, .88, .95\}$.  TrICP and FICP return similar values of \rmsd and \frmsd on average while also determining reasonable values for $f$.  However, FICP is about $6\times$ faster than TrICP using the golden ratio search.  

\begin{table}[h!!!t]
\begin{tabular}{|r|c|c|c|c|c|c|c|c|}
\hline
Algorithm & $p_I$ & time (s) & \# iterations & \rmsd & \frmsd & f \\
\hline 
ICP &  .75 & 0.335 & 24.5 & 9.454 & 9.454 & 1.000 \\
TrICP & .75 & 1.356 & 117.9 & 0.217 & 0.541 & 0.744 \\
FICP & .75 & 0.178 & 13.6 &  0.178 & 0.424 & 0.749 \\
\hline
ICP &  .88 & 0.21 & 21.4 & 4.079 & 4.079 & 1.000 \\
TrICP & .88 & 1.032 & 107.5 & 0.218 & 0.364 & 0.784 \\
FICP & .88 &  0.136 &  12.3 &  0.175 & 0.258 & 0.878 \\
\hline
ICP & .95 & 0.137 & 15.9 & 1.338 & 1.338 & 1.000 \\
TrICP &  .95 & 0.913 & 102.4 & 0.197 & 0.261 & 0.904 \\
FICP & .95 & 0.123 & 12.0 & 0.175 & 0.205 & 0.949 \\
\hline
\end{tabular}
\caption{SQUID data with \emph{Occlusion} outliers, rotated $5^\circ$}
\label{tbl:SQUID-O5}
\end{table}

\begin{table}[htb]
\begin{tabular}{|r|c|c|c|c|c|c|c|c|}
\hline
Algorithm & $p_I$ &time (s) & \# iterations & \rmsd & \frmsd & f \\
\hline 
ICP &  .75 & 0.263 & 28.9 & 1.074 & 1.074 & 1.000 \\
TrICP & .75 & 1.103 & 114.8 & 0.213 & 0.404 & 0.803 \\
FICP & .75 & 0.191 & 18.1 &  0.231 & 0.402 & 0.810 \\
\hline
ICP &  .88 & 0.215 & 24.4 & 0.829 & 0.829 & 1.000 \\
TrICP & .88 & 1.065 & 112.8 & 0.213 & 0.335 & 0.827 \\
FICP & .88 & 0.148 & 14.2 & 0.178 & 0.241 & 0.903 \\
\hline
ICP &  .95 & 0.168 & 19.6 & 0.569 & 0.569 & 1.000 \\
TrICP & .95 & 1.020 & 111.6 & 0.203 & 0.281 & 0.900 \\
FICP & .95 & 0.138 & 13.3 & 0.174 & 0.197 & 0.959 \\
\hline
\end{tabular}
\caption{SQUID data with \emph{Deformation} outliers, rotated $5^\circ$}
\label{tbl:SQUID-D5}
\end{table}

\begin{table}[htb]
\begin{tabular}{|r|c|c|c|c|c|c|c|c|}
\hline
Algorithm & $p_I$ & time (s) & \# iterations & \rmsd & \frmsd & f \\
\hline 
ICP &.75 &  0.461 & 26.7 & 5.820 & 5.820 & 1.000 \\
TrICP &.75 & 1.578 & 92.9 & 0.176 & 0.399 & 0.768 \\
FICP &.75 &  0.264 & 13.7 &  0.175 & 0.388 & 0.766 \\
\hline
ICP &  .88 & 0.286 & 23.7 & 4.061 & 4.061 & 1.000 \\
TrICP & .88 & 1.351 & 108.0 & 0.202 & 0.309 & 0.831 \\
FICP & .88 & 0.183 & 13.1 & 0.172 & 0.246 & 0.888 \\
\hline
ICP &  .95 & 0.192 & 19.5 & 2.626 & 2.626 & 1.000 \\
TrICP & .95 & 1.135 & 108.3 & 0.205 & 0.295 & 0.893 \\
FICP & .95 & 0.148 & 12.6 & 0.171 & 0.197 & 0.953 \\
\hline
\end{tabular}
\caption{SQUID data with \emph{New Data} outliers, rotated $5^\circ$}
\label{tbl:SQUID-N5}
\end{table}

The $f$ values when deformation outliers are introduced are noticeably larger than $p_I$ because some of the shifted points happen to lie very near model points when the two data sets are properly aligned.  These points might as well be inliers.  This phenomenon is less common for the other types of synthetically generated outliers.  

We also ran the same experiments with the same algorithms on the bunny (35,947 points) and happy Buddha (144,647 points) data from the Stanford 3D scanning repository.  We report the results on the bunny data set in Table \ref{tbl:bunny-D5} and for the happy Buddha data set in Table \ref{tbl:happy-D5} where deformation outliers are applied to $D$ and then $D$ is randomly rotated by $5^\circ$.  The numbers are the the results of averages over $10$ random rotations.

\begin{table}[h!!!t]
\begin{tabular}{|r|c|c|c|c|c|c|c|}
\hline
Algorithm & $p_I$ & time (s) & \# iterations & \rmsd & \frmsd & f \\
\hline 
ICP & .75 & 60.1  & 78.8 & 0.66682 & 0.66682 & 1.000\\
TrICP & .75 & 136.5 & 172.2 & 0.00523 & 0.01239 & 0.750\\
FICP &  .75 & 16.5 & 17.3 & 0.00522 & 0.01237 & 0.750 \\
\hline
ICP &  .88 & 29.6 & 48.0 & 0.45303 & 0.45303 & 1.000 \\
TrICP & .88 & 147.1 & 224.3 & 0.00522 & 0.00767 & 0.880 \\
FICP & .88 & 13.7 & 15.9 &  0.00522 & 0.00767 & 0.880 \\
\hline
ICP & .95 & 13.8 & 31.3 & 0.37207 & 0.37207 & 1.000\\
TrICP & .95 & 77.6 & 162.8 & 0.00523 & 0.00610 & 0.950 \\
FICP & .95 & 8.0 & 14.2 & 0.00523 & 0.00610 & 0.950\\
\hline
\end{tabular}
\caption{bunny with \emph{Deformation} outliers, rotated $5^\circ$}
\label{tbl:bunny-D5}
\end{table}

\begin{table}[h!!!t]
\begin{tabular}{|r|c|c|c|c|c|c|c|c|}
\hline
Algorithm & $p_I$ & time (s) & \# iterations & \rmsd & \frmsd & f \\
\hline 
ICP & .75 & 430.8 & 66.6 & 0.56145 & 0.56146 & 1.000 \\
TrICP & .75 & 727.3 & 101.0 & 0.00123 & 0.00291 & 0.750 \\
FICP & .75 & 139.7 & 15.6 & 0.00119 & 0.00282 & 0.750 \\
\hline 
ICP & .88 &  109.2 & 28.2 & 0.29745 & 0.29745 & 1.000 \\
TrICP & .88 & 485.4 & 120.5 & 0.00120 & 0.00177 & 0.880 \\
FICP & .88 & 81.4 & 15.7 &  0.00119 & 0.00174 & 0.880 \\
\hline 
ICP & .95 & 126.1 & 45.4 & 0.29351 & 0.29351 & 1.000 \\
TrICP & .95 & 405.2 & 123.7 & 0.00120 & 0.00141 & 0.950 \\
FICP & .95 & 66.3 & 14.6 & 0.00119 & 0.00139 & 0.950 \\
\hline
\end{tabular}
\caption{Buddha with \emph{Deformation} outliers, rotated $5^\circ$}
\label{tbl:happy-D5}
\end{table}

\begin{figure}[htb]
  \centering 
  \includegraphics[width=.25\linewidth]{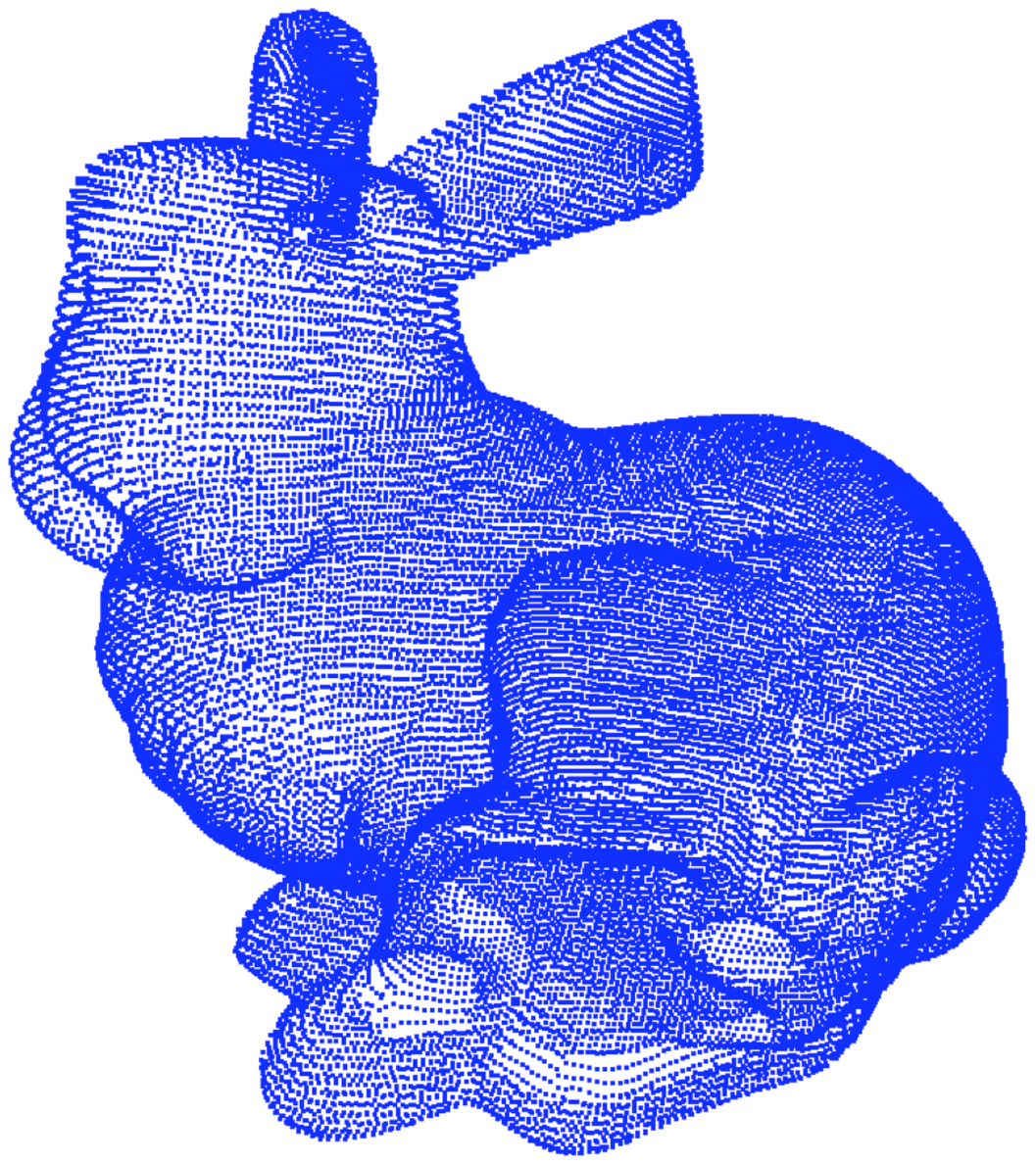}
  \hspace{1.3in}
  \includegraphics[width=.25\linewidth]{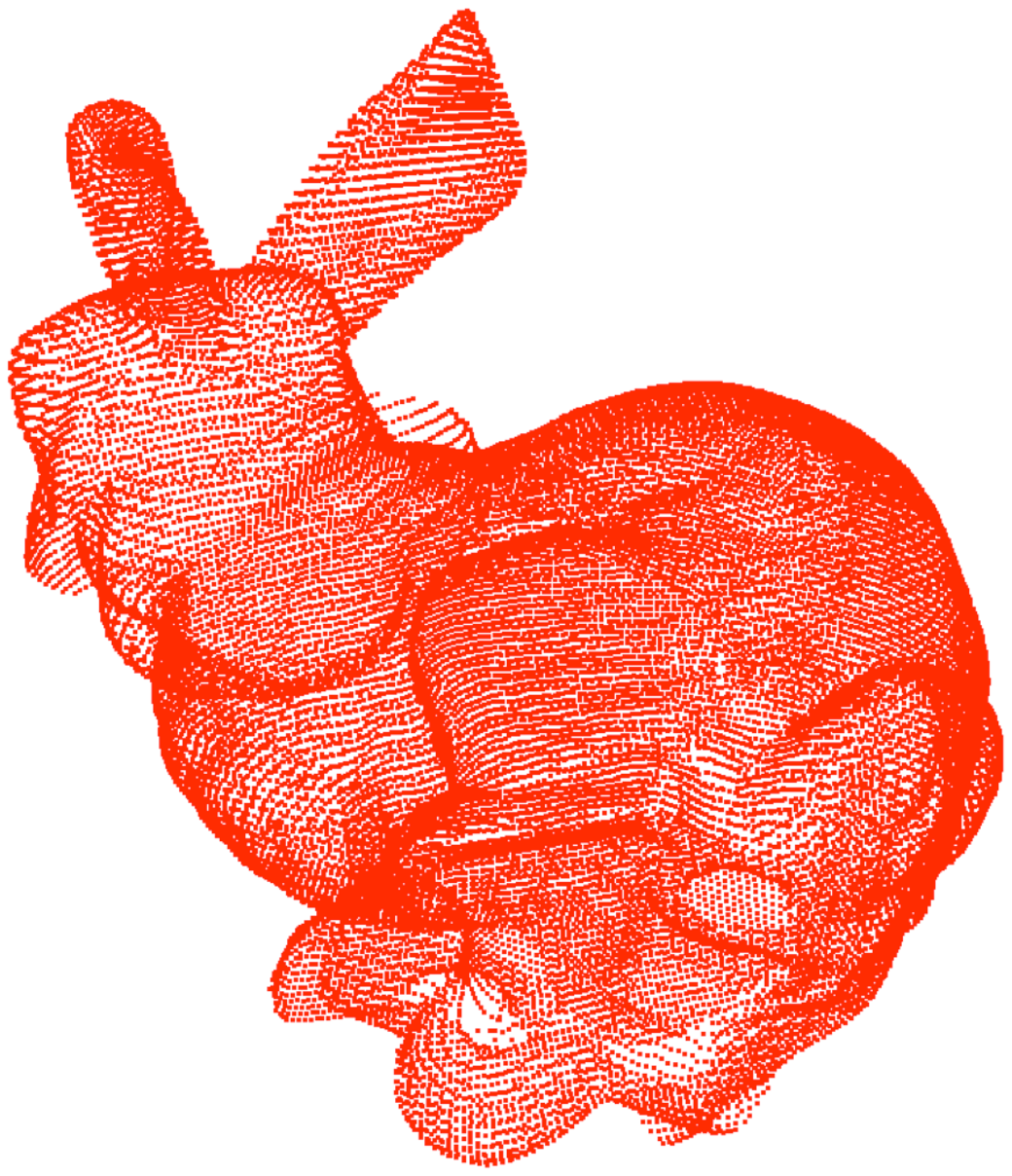}
  \\
  \includegraphics[width=.45\linewidth]{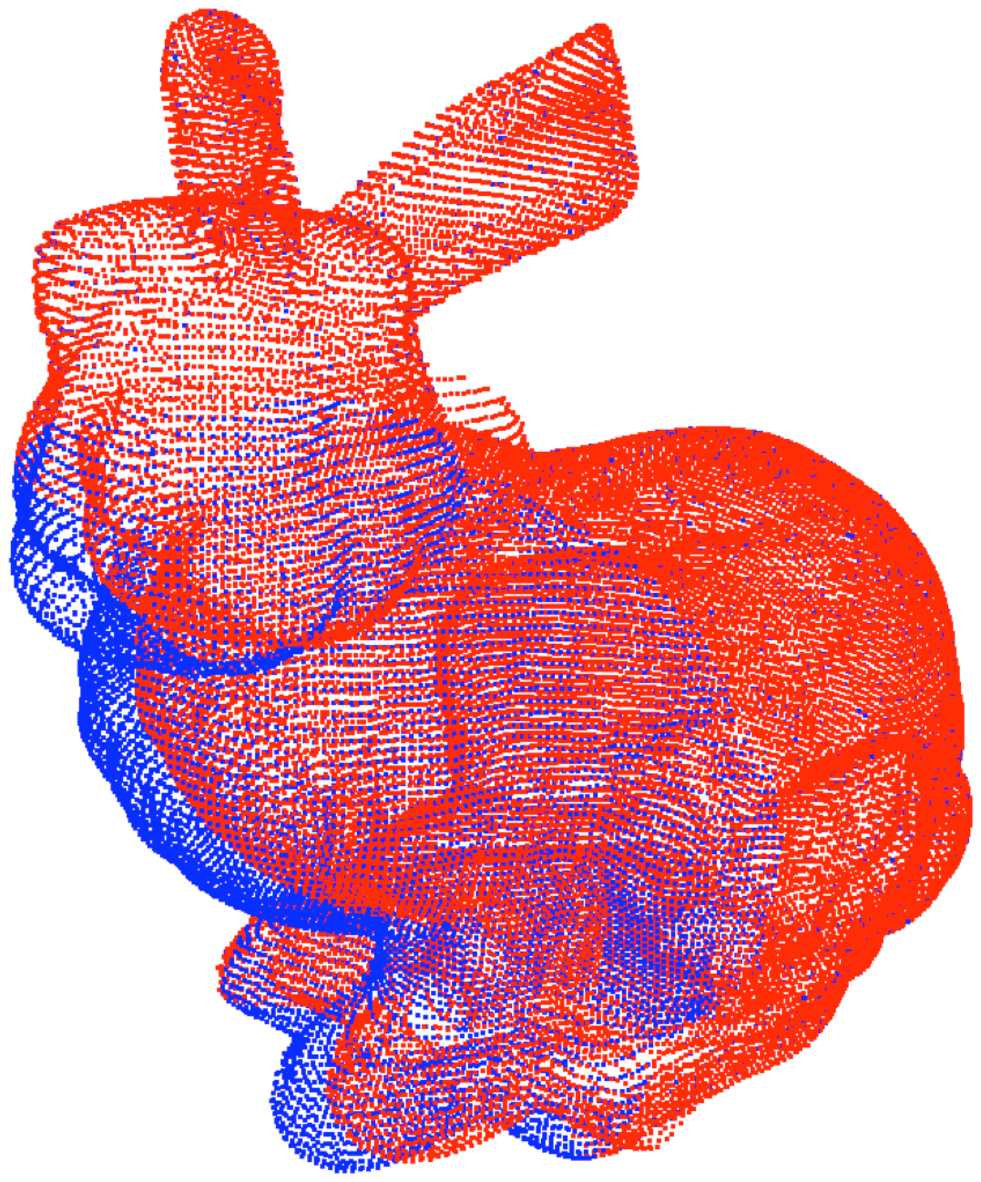}
  \includegraphics[width=.5\linewidth]{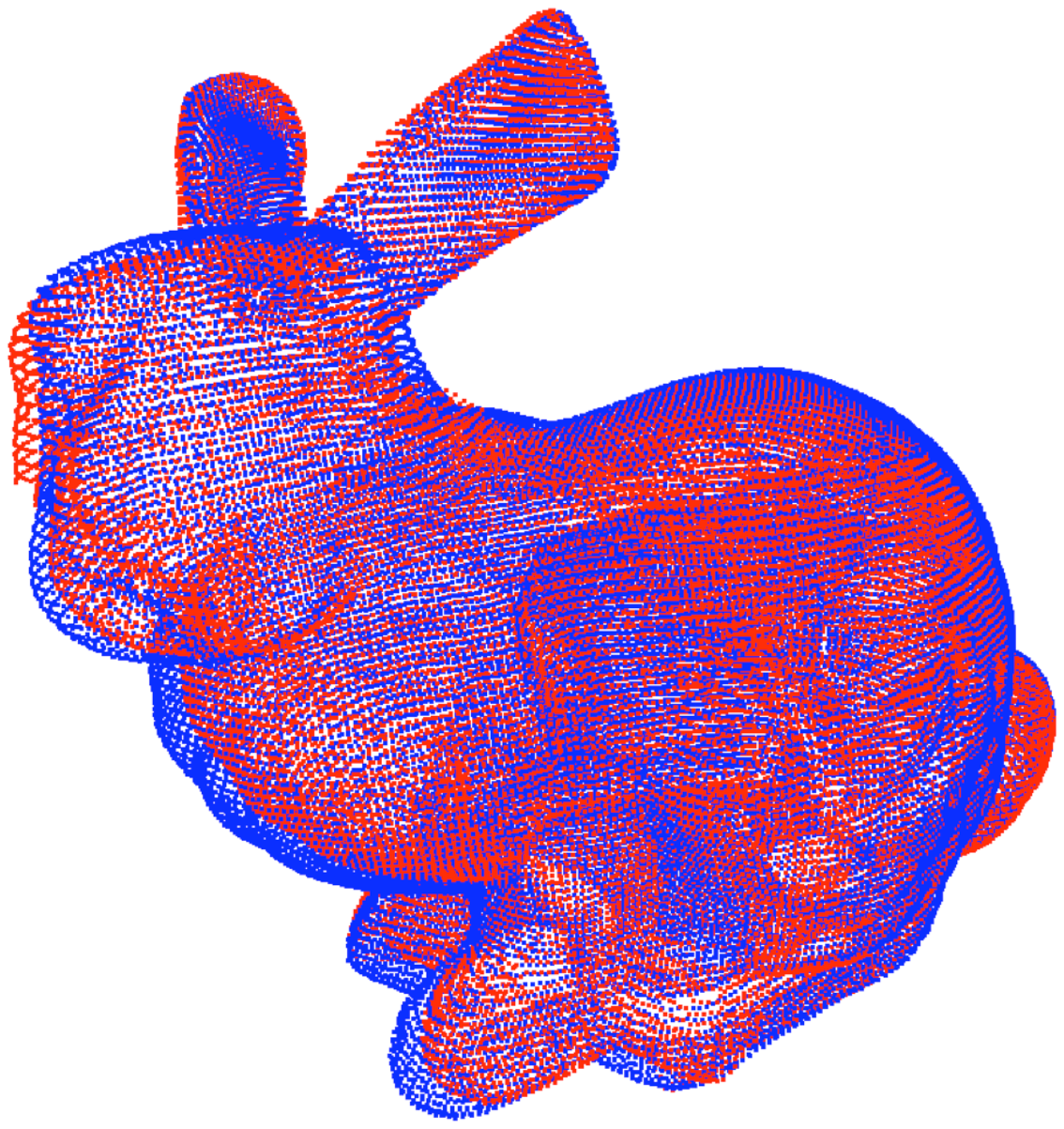}
  \caption{\label{fig:bunny+D}
  	Stanford bunny with $M$ in blue (top left) and $D$ in red (top right) with \emph{Deformation} noise with $p_I = .75$.  Registered using FICP (bottom left) and ICP (bottom right).}
\end{figure}

Observe in Figure \ref{fig:bunny+D} how in the alignment of the bunny data set, the non-deformed points (red points on back side, blue points are not visible because they lie exactly behind the red points) are aligned almost exactly by the FICP algorithm while the deformed points (shifted from visible blue points in front) are ignored.  Such an alignment allows one to easily identify the portion of the data which has been deformed, and by how much it has been deformed.  Without a proper registration to the model the unaligned points have no point of comparison to gauge their deformation.  The alignment is skewed when ICP is used and it is not helpful in determining which points are deformed.

\subsection{Funnel of Convergence}
We calculate the percentage of cases from the SQUID data set that converge to an \frmsd value within $.01$ and $f$ value within $.01$ of the alignment between the same sets with no initial rotation.  Table \ref{tbl:rotate} shows the results when New Data outliers with $p_I = .88$ are added to the data set $D$.  The results for the other types of noise are simlar.  For 3D data sets we chose $\sigma$ proportionally smaller, so these convergence rates are all slightly larger.
Note that FICP with $\lambda = 3$ performs much better than when $\lambda = 1.3$.  

\begin{table}[h!!!t]
\begin{tabular}{|r|c|c|c|c|c|c|c|}
\hline
Algorithm & $\lambda$ & $5^\circ$ & $10^\circ$ & $25^\circ$ & $50^\circ$ \\
\hline 
ICP & - & 0.999 & 0.997 & 0.994 & 0.962 \\
TrICP & 3 & 0.875 & 0.870 & 0.853 & 0.816 \\
FICP & 3 & 0.952 &  0.945 & 0.909 & 0.875 \\
FICP & 1.3 & 0.857 & 0.473 & 0.141 & 0.060 \\
\hline
\end{tabular}
\caption{Percentage of SQUID data sets converging per initial rotation.}
\label{tbl:rotate}
\end{table}

ICP has a larger radius of convergence than FICP, because it searches a much smaller parameter space.  FICP has a larger radius of convergence than TrICP even though they search the same parameter space.

\subsection{Validating $\lambda$}
\label{sec:val-lambda}
We empirically justify that \frmsd is not sensitive to the choice of $\lambda$.  We run FICP with $\lambda$ set to $\{1, 1.3, 2, 3, 4, 5\}$.  We plot the averaged results on the SQUID data set when Occlusion noise is added with $p_I = .75$ and $D$ is initially rotated $0^\circ$ and $5^\circ$ in Table \ref{tbl:lambda0} and Table \ref{tbl:lambda5}, respectively.  Altering $\lambda$ does not dramatically affect the converged solution, but can affect the radius of convergence.  The output is similar for different types of noise.  On 3D data, FICP performs slightly better than 2D data for smaller $\lambda$.

\begin{table}[h!!!t]
\begin{tabular}{|r|r|c|c|c|c|c|c|c|}
\hline
Algorithm & $\lambda$ & time (s) & \# iterations & \rmsd & \frmsd & f \\
\hline 
FICP & 1 & 0.142 & 10.38 & 0.158 & 0.225 & 0.701\\
FICP & 1.3 &  0.069 & 3.81 & 0.170 & 0.248 & 0.749\\
FICP & 2 & 0.059 & 3.06 & 0.170  & 0.303 & 0.750\\
FICP & 3 &  0.061 & 3.17 & 0.170  & 0.404 & 0.750\\
FICP & 4 & 0.062 & 3.21 & 0.171  & 0.538 & 0.751\\
FICP & 5 & 0.063 & 3.30 & 0.172  & 0.717 & 0.751\\
\hline
\end{tabular}
\caption{FICP for different values of $\lambda$ with $D$ rotated $0^\circ$.}
\label{tbl:lambda0}
\end{table}

\begin{table}[h!!!t]
\begin{tabular}{|r|r|c|c|c|c|c|c|c|}
\hline
Algorithm & $\lambda$ & time (s) & \# iterations & \rmsd & \frmsd & f \\
\hline 
FICP & 1 & 0.733 & 37.23 & 0.298 & 1.503 & 0.274\\
FICP & 1.3 &  0.488 & 36.44 & 0.219 & 0.563 & 0.660\\
FICP & 2 & 0.244 & 17.00 & 0.176  & 0.329 & 0.740\\
FICP & 3 &  0.198 & 13.59 & 0.178  & 0.424 & 0.749\\
FICP & 4 & 0.194 & 13.28 & 0.184  & 0.570 & 0.751\\
FICP & 5 & 0.200 & 13.66 & 0.299  & 0.875 & 0.756\\
\hline
\end{tabular}
\caption{FICP for different values of $\lambda$ with $D$ rotated $5^\circ$.}
\label{tbl:lambda5}
\end{table}

\subsection{Aligning Scanned Model Data}
Finally, we perform experiments aligning real scanned range maps from 3D models.  We consider aligning two scans from the Stanford 3D scanning repository of the dragon model.  We take scans from the \emph{dragonStandRight} data set and we align consecutive scans ($24^\circ$ apart), as seen in Table \ref{tbl:dragon24}, and next-to-consecutive scans ($48^\circ$ apart), as seen in Table \ref{tbl:dragon48}.  We first rotate the later scan by $24^\circ$ or $48^\circ$ to bring the scans into the approximately correct alignment.  We then align them with ICP, TrICP, and FICP.

\begin{table}[h!!!t]
\begin{tabular}{|r|c|c|c|c|c|c|c|}
\hline
Algorithm & angle1 & angle2 & time (s) & \# iterations & \rmsd & \frmsd & f \\
\hline 
ICP & 336 & 0 & 40.88 & 62 & 0.001150 & 0.001150 & 1.000\\
TrICP & 336 & 0 & 316.33 & 535 & 0.000193 & 0.000303 & 0.860\\
FICP & 336 & 0 & 35.03 & 53 & 0.000193 & 0.000303 & 0.862\\
\hline
ICP & 0 & 24 & 22.55 & 44 & 0.001059 & 0.001059 & 1.000\\
TrICP & 0 & 24 & 337.09 & 709 & 0.000186 & 0.000251 & 0.904\\
FICP & 0 & 24 & 28.22 & 54 & 0.000186 & 0.000251 & 0.905\\
\hline
ICP & 24 & 48 & 36.70 & 49 & 0.003207 & 0.003207 & 1.000\\
TrICP & 24 & 48 & 346.56 & 761 & 0.000197 & 0.000292 & 0.877\\
FICP & 24 & 48 & 42.41 & 90 & 0.000198 & 0.000291 & 0.879\\
\hline
ICP & 48 & 72 & 80.37 & 50 & 0.004003 & 0.004003 & 1.000\\
TrICP & 48 & 72 & 771.72 & 519 & 0.000206 & 0.000894 & 0.613\\
FICP & 48 & 72 & 84.39 & 54 & 0.000208 & 0.000894 & 0.615\\
\hline
ICP & 72 & 96 & 229.48 & 66 & 0.007456 & 0.007456 & 1.000\\
TrICP & 72 & 96 & 915.01 & 485 & 0.000204 & 0.000786 & 0.638\\
FICP & 72 & 96 & 140.79 & 69 & 0.000205 & 0.000786 & 0.639\\
\hline
ICP & 96 & 120 & 132.56 & 47 & 0.003806 & 0.003806 & 1.000\\
TrICP & 96 & 120 & 1444.58 & 506 & 0.000190 & 0.000926 & 0.590\\
FICP & 96 & 120 & 206.06 & 66 & 0.000190 & 0.000926 & 0.589\\
\hline
ICP & 120 & 144 & 194.36 & 60 & 0.003915 & 0.003915 & 1.000\\
TrICP & 120 & 144 & 2066.43 & 836 & 0.000192 & 0.000453 & 0.751\\
FICP & 120 & 144 & 182.97 & 70 & 0.000192 & 0.000453 & 0.752\\
\hline
ICP & 144 & 168 & 59.90 & 67 & 0.001185 & 0.001185 & 1.000\\
TrICP & 144 & 168 & 525.75 & 633 & 0.000189 & 0.000296 & 0.862\\
FICP & 144 & 168 & 74.77 & 84 & 0.000189 & 0.000296 & 0.862\\
\hline
ICP & 168 & 192 & 46.56 & 64 & 0.000605 & 0.000605 & 1.000\\
TrICP & 168 & 192 & 580.48 & 967 & 0.000188 & 0.000251 & 0.908\\
FICP & 168 & 192 & 61.57 & 88 & 0.000188 & 0.000251 & 0.909\\
\hline
ICP & 192 & 216 & 101.19 & 74 & 0.002759 & 0.002759 & 1.000\\
TrICP & 192 & 216 & 1049.67 & 1297 & 0.000177 & 0.000247 & 0.895\\
FICP & 192 & 216 & 82.98 & 91 & 0.000176 & 0.000246 & 0.895\\
\hline
ICP & 216 & 240 & 41.64 & 79 & 0.000860 & 0.000860 & 1.000\\
TrICP & 216 & 240 & 459.49 & 758 & 0.000194 & 0.000317 & 0.849\\
FICP & 216 & 240 & 46.33 & 73 & 0.000195 & 0.000317 & 0.845\\
\hline
ICP & 240 & 264 & 85.09 & 52 & 0.004253 & 0.004253 & 1.000\\
TrICP & 240 & 264 & 687.99 & 577 & 0.000202 & 0.000442 & 0.770\\
FICP & 240 & 264 & 87.90 & 73 & 0.000202 & 0.000441 & 0.771\\
\hline
ICP & 264 & 288 & 568.15 & 100 & 0.011210 & 0.011210 & 1.000\\
TrICP & 264 & 288 & 3486.41 & 627 & 0.000181 & 0.001517 & 0.492\\
FICP & 264 & 288 & 342.03 & 57 & 0.000185 & 0.001511 & 0.496\\
\hline
ICP & 288 & 312 & 142.53 & 45 & 0.003097 & 0.003097 & 1.000\\
TrICP & 288 & 312 & 1559.13 & 528 & 0.000195 & 0.001032 & 0.574\\
FICP & 288 & 312 & 170.86 & 53 & 0.000207 & 0.001056 & 0.581\\
\hline
ICP & 312 & 336 & 42.65 & 43 & 0.000967 & 0.000967 & 1.000\\
TrICP & 312 & 336 & 640.96 & 713 & 0.000197 & 0.000338 & 0.835\\
FICP & 312 & 336 & 52.42 & 49 & 0.000197 & 0.000338 & 0.836\\
\hline
\end{tabular}
\caption{Alignment of dragon scans off by $24^\circ$ with ICP, TrICP, and FICP.}
\label{tbl:dragon24}
\end{table}

\begin{table}[h!!!t]
\begin{tabular}{|r|c|c|c|c|c|c|c|}
\hline
Algorithm & angle1 & angle2 & time (s) & \# iterations & \rmsd & \frmsd & f \\
\hline 
ICP & 312 & 0 & 112.69 & 50 & 0.002191 & 0.002191 & 1.000 \\
TrICP & 312 & 0 & 1681.35 & 803 & 0.000221 & 0.000759 & 0.663 \\
FICP & 312 & 0 & 188.44 & 84 & 0.000217 & 0.000759 & 0.659 \\
\hline
ICP & 336 & 24 & 83.43 & 71 & 0.002067 & 0.002067 & 1.000 \\
TrICP & 336 & 24 & 617.85 & 629 & 0.000207 & 0.000507 & 0.742 \\
FICP & 336 & 24 & 88.85 & 87 & 0.000208 & 0.000507 & 0.743 \\
\hline
ICP & 0 & 48 & 54.37 & 53 & 0.003417 & 0.003417 & 1.000 \\
TrICP & 0 & 48 & 804.91 & 1087 & 0.000205 & 0.000480 & 0.753 \\
FICP & 0 & 48 & 77.74 & 101 & 0.000206 & 0.000479 & 0.755 \\
\hline
ICP & 24 & 72 & 164.46 & 65 & 0.005940 & 0.005940 & 1.000 \\
TrICP & 24 & 72 & 1384.2 & 680 & 0.004822 & 0.005814 & 0.940 \\
FICP & 24 & 72 & 223.19 & 86 & 0.005788 & 0.005896 & 0.994 \\
\hline
ICP & 48 & 96 & 386.95 & 156 & 0.005776 & 0.005776 & 1.000 \\
TrICP & 48 & 96 & 3167.94 & 1273 & 0.005601 & 0.005756 & 0.991 \\
FICP & 48 & 96 & 439.68 & 173 & 0.005599 & 0.005756 & 0.991 \\
\hline
ICP & 72 & 120 & 763.38 & 76 & 0.012262 & 0.012262 & 1.000 \\
TrICP & 72 & 120 & 2929.36 & 311 & 0.000525 & 0.008209 & 0.400 \\
FICP & 72 & 120 & 721.8 & 67 & 0.010804 & 0.012084 & 0.963 \\
\hline
ICP & 96 & 144 & 338.17 & 54 & 0.006428 & 0.006428 & 1.000 \\
TrICP & 96 & 144 & 2512.57 & 400 & 0.000241 & 0.003770 & 0.400 \\
FICP & 96 & 144 & 480.89 & 77 & 0.002191 & 0.005132 & 0.753 \\
\hline
ICP & 120 & 168 & 495.54 & 91 & 0.004723 & 0.004723 & 1.000 \\
TrICP & 120 & 168 & 3824.02 & 838 & 0.000209 & 0.001108 & 0.573 \\
FICP & 120 & 168 & 525.3 & 110 & 0.000208 & 0.001108 & 0.573 \\
\hline
ICP & 144 & 192 & 156.29 & 77 & 0.001936 & 0.001936 & 1.000 \\
TrICP & 144 & 192 & 2167.88 & 1415 & 0.000210 & 0.000574 & 0.715 \\
FICP & 144 & 192 & 243.71 & 149 & 0.000210 & 0.000574 & 0.716 \\
\hline
ICP & 168 & 216 & 205.34 & 88 & 0.003037 & 0.003037 & 1.000 \\
TrICP & 168 & 216 & 2830.94 & 1428 & 0.000197 & 0.000396 & 0.793 \\
FICP & 168 & 216 & 297.73 & 136 & 0.000198 & 0.000396 & 0.794 \\
\hline
ICP & 192 & 240 & 271.59 & 115 & 0.004515 & 0.004515 & 1.000 \\
TrICP & 192 & 240 & 2459.96 & 762 & 0.000193 & 0.000720 & 0.645 \\
FICP & 192 & 240 & 225.21 & 114 & 0.000194 & 0.000720 & 0.646 \\
\hline
ICP & 216 & 264 & 344.86 & 138 & 0.005295 & 0.005295 & 1.000 \\
TrICP & 216 & 264 & 2488.24 & 664 & 0.002994 & 0.006536 & 0.771 \\
FICP & 216 & 264 & 491.86 & 212 & 0.000238 & 0.001304 & 0.568 \\
\hline
ICP & 240 & 288 & 181.26 & 49 & 0.006412 & 0.006412 & 1.000 \\
TrICP & 240 & 288 & 2488.24 & 731 & 0.005687 & 0.006262 & 0.968 \\
FICP & 240 & 288 & 168.24 & 47 & 0.005719 & 0.006262 & 0.970 \\
\hline
ICP & 264 & 312 & 1093.75 & 79 & 0.013483 & 0.013483 & 1.000 \\
TrICP & 264 & 312 & 4417.08 & 675 & 0.013477 & 0.013483 & 1.000 \\
FICP & 264 & 312 & 1115.72 & 79 & 0.013483 & 0.013483 & 1.000 \\
\hline
ICP & 288 & 336 & 193.36 & 39 & 0.003856 & 0.003856 & 1.000 \\
TrICP & 288 & 336 & 2324.38 & 511 & 0.000236 & 0.003080 & 0.425 \\
FICP & 288 & 336 & 295.88 & 61 & 0.002842 & 0.003617 & 0.923 \\
\hline
\end{tabular}
\caption{Alignment of dragon scans off by $48^\circ$ with ICP, TrICP, and FICP.}
\label{tbl:dragon48}
\end{table}

For most alignments both FICP and TrICP realize an alignment with a much lower \frmsd value than ICP.  And occasionally, FICP noticeably outperforms TrICP in this regard as well.  FICP is usually about as fast as ICP, and is consistently about $5$ to $10$ times faster than TrICP.  Notice how as the solution for FICP has $f$ approach $1$, then FICP gracefully approaches the result of ICP with very little noticeable overhead.  

\begin{figure}[htb]
  \centering 
  \includegraphics[width=.48\linewidth]{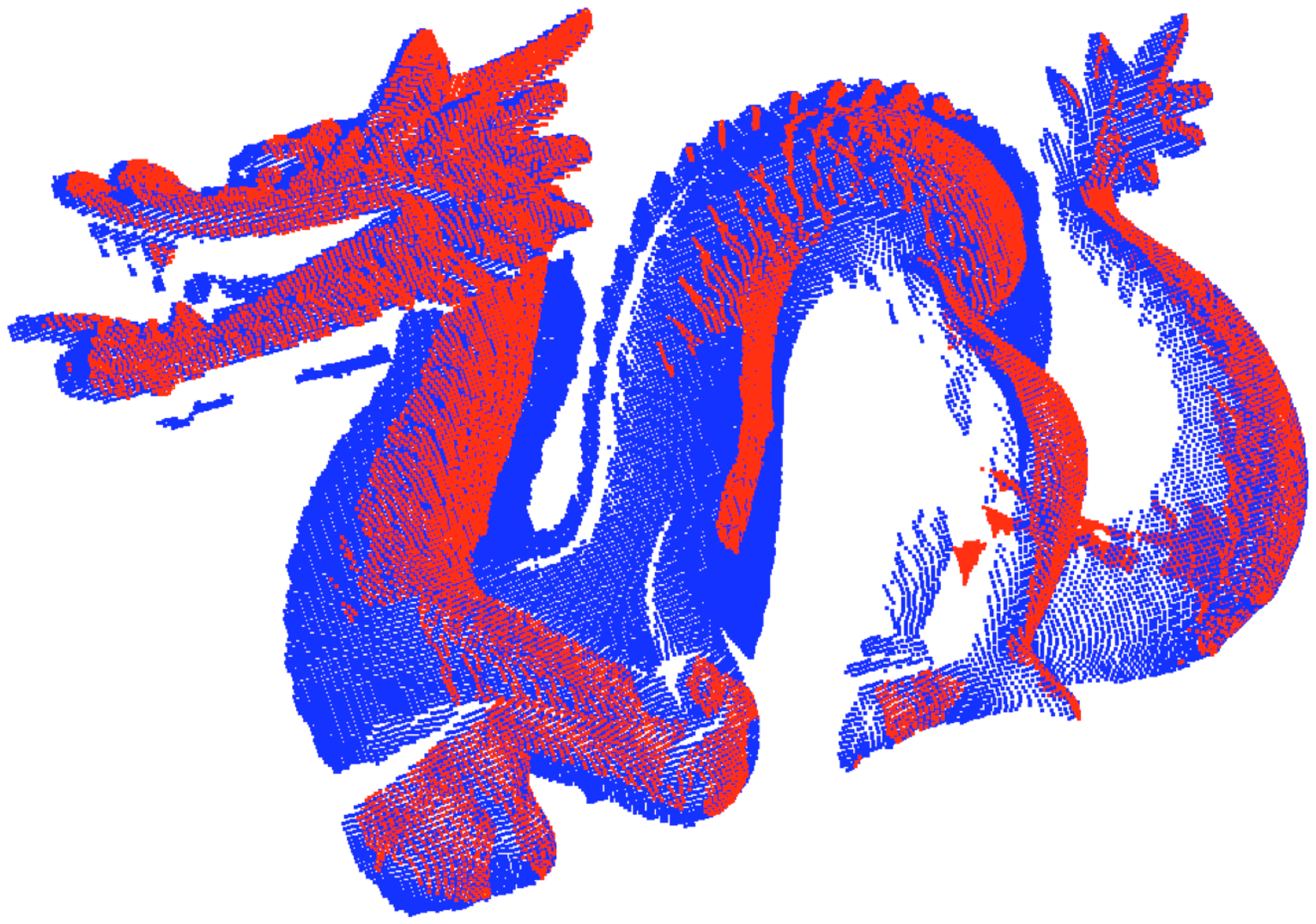}
  \includegraphics[width=.48\linewidth]{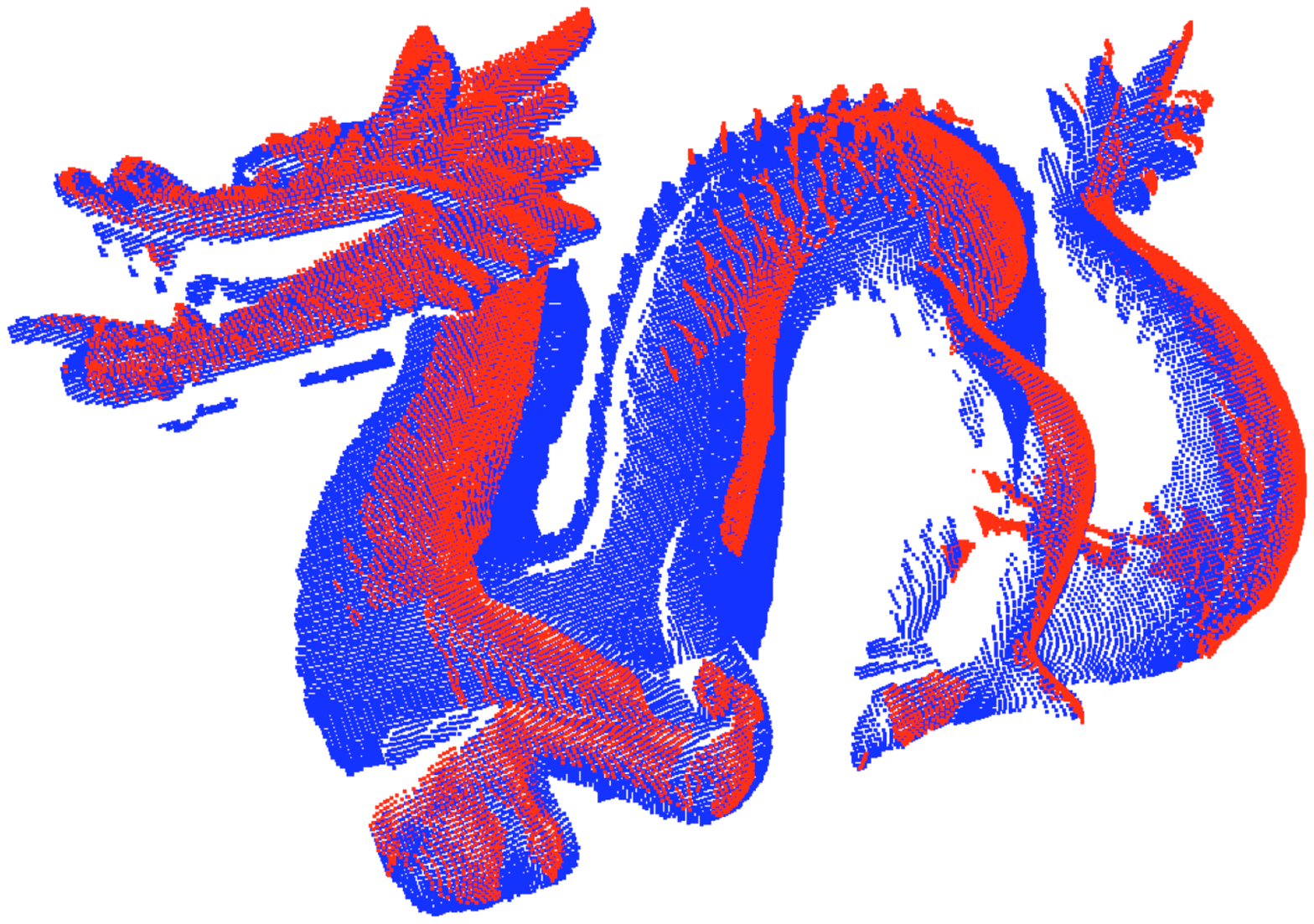}
  \\
  \includegraphics[width=.48\linewidth]{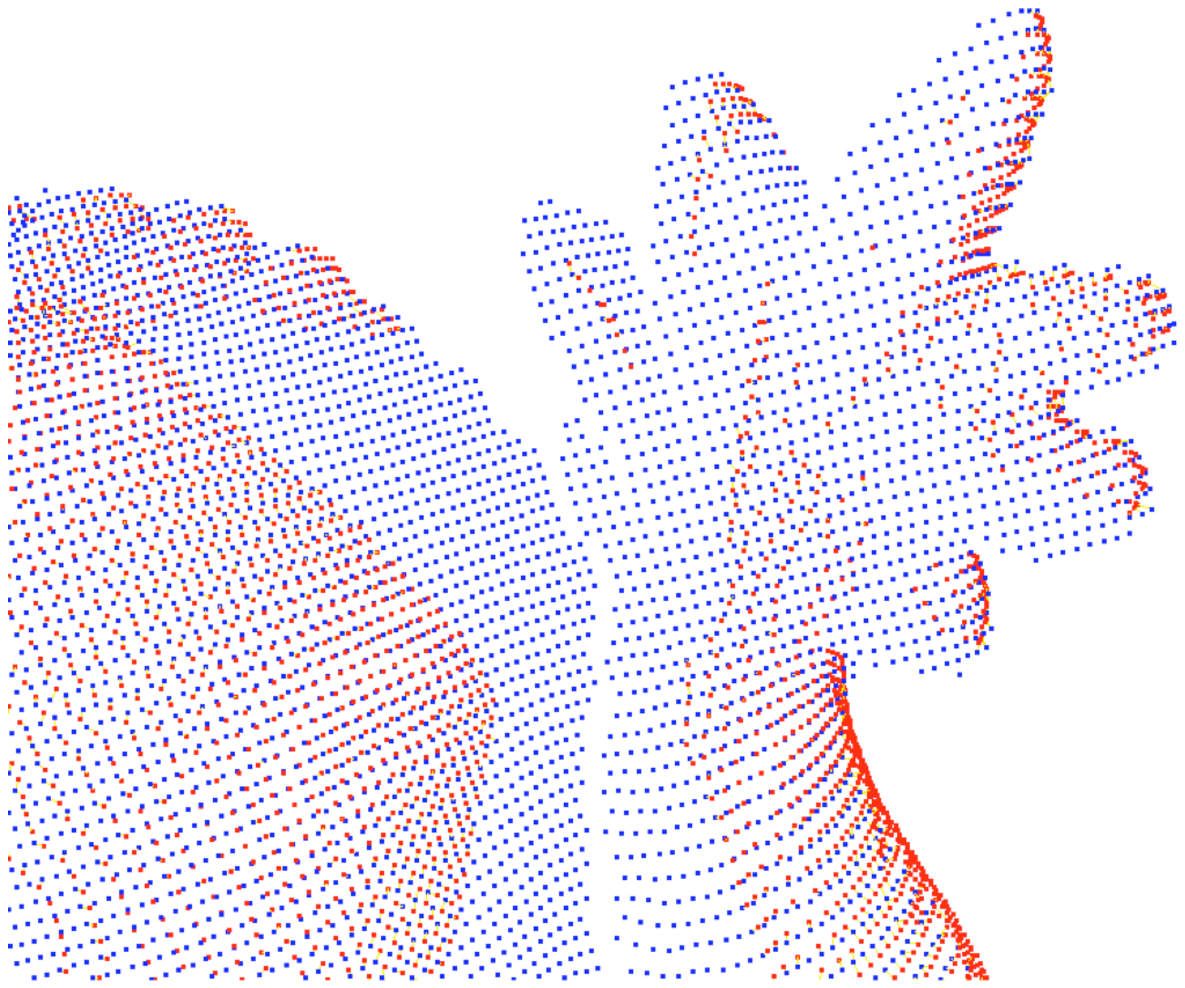}
  \includegraphics[width=.48\linewidth]{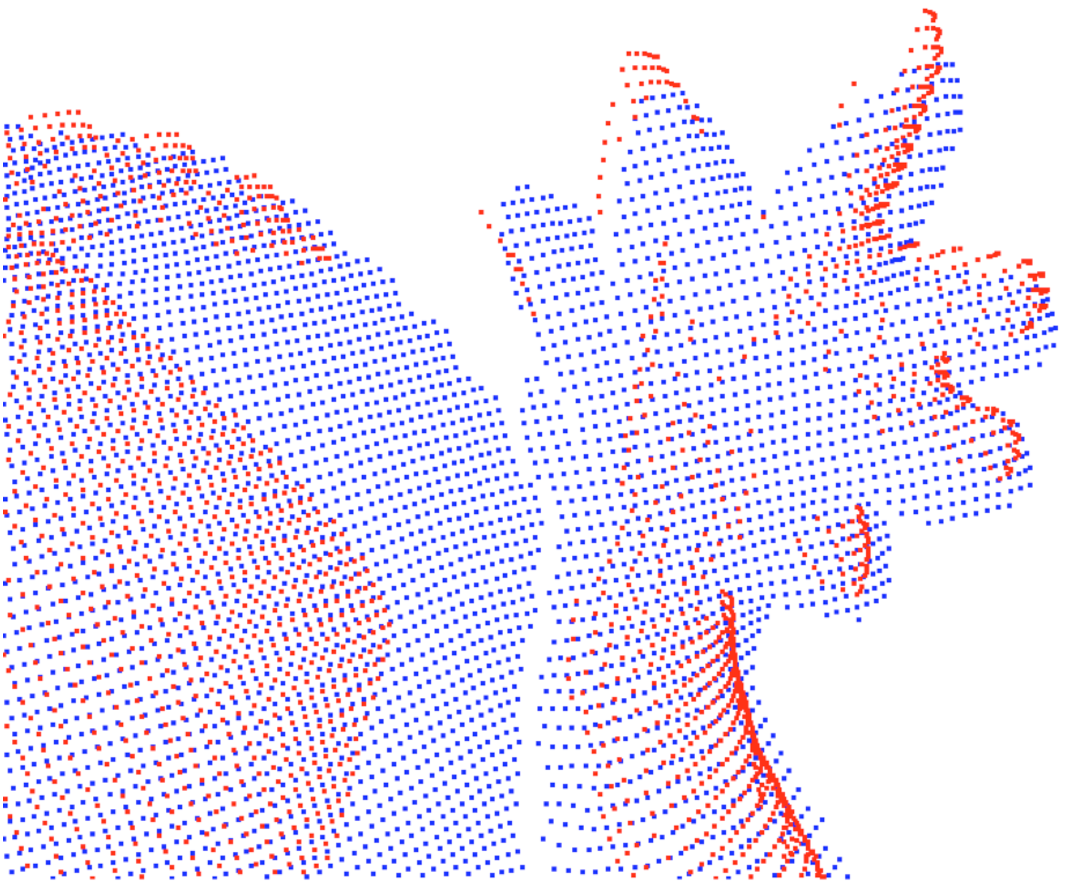}  
  \caption{\label{fig:dragon}
  	Stanford dragon scans at $0^\circ$ and $48^\circ$ with $M$ in blue and $D$ in red, registered using FICP (top left) and ICP (top right).  Zoomed images of the alignment around dragon's tail with FICP (bottom left) and ICP (bottom right) are shown to demonstrate the skew in the alignment due to ICP.}
\end{figure}

Figure \ref{fig:dragon} shows the alignment of the scan at $0^\circ$ aligned with the scan at $48^\circ$ using ICP and FICP.  Notice how when the scans are aligned with ICP, the points in the dragon's tail are slightly misaligned, whereas with FICP, the alignment is much better.  This is confirmed in Table \ref{tbl:dragon48}.

\section{Conclusion}
In considering the common problem of aligning two points sets under a set of transformations, we specifically handle the problem of outliers.  We formalize the distance measure \frmsd (a generalization of \rmsd), and we provide an algorithm, FICP, to efficiently solve for a local minimum in this distance under a set of transformations, all possible matchings, and the set of outliers.  We prove that FICP converges to a local minimum, and that under reasonable assumptions on the data, this minimum chooses a set of inliers such that each point selected is more likely to be an inlier than an outlier, and each point not selected is more likely to be an outlier than an inlier.  On a variety of synthetic data and real scanned range maps we show that FICP compares favorably to alternative algorithms which are guaranteed to converge\textemdash ICP and TrICP.  Because this algorithm is a very simple modification of the quite popular ICP algorithm and it is compatible with most other recent improvements, we expect that these ideas will be integrated into many modern systems.  

\bibliographystyle{plain}
\bibliography{FICPbib}

\end{document}